\documentclass[lineno]{jfm}

\usepackage{graphicx}
\usepackage{gensymb}
\usepackage{subfigure}
\usepackage{float}
\usepackage{epstopdf}
\usepackage{epsfig}
\usepackage[figuresright]{rotating}
\usepackage{adjustbox}
\usepackage{makecell}
\usepackage{bm}
\usepackage{newtxtext}
\usepackage{newtxmath}
\usepackage{natbib}
\usepackage{hyperref}
\hypersetup{
    colorlinks = true,
    urlcolor   = red,
    citecolor  = blue,
}

\newcommand{\RomanNumeralCaps}[1]
\linenumbers

\title{On the shock wave boundary layer interaction in slightly-rarefied gas  }

\author{
	Hualin Liu\aff{1}, Qi Li\aff{1}, Weifang Chen\aff{2}
	\and Lei Wu\aff{1}
 \corresp{\email{wul@sustech.edu.cn}}
}

\affiliation{\aff{1} Department of Mechanics and Aerospace Engineering, Southern University of Science and Technology, Shenzhen 518055, China 
\aff{2} School of Aeronautics and Astronautics, Zhejiang University,  Hangzhou 310027, China}
\begin{document}
\maketitle

\begin{abstract}
The shock wave and boundary layer interaction (SWBLI) plays an important role in the design of hypersonic vehicles. However, discrepancies between the numerical results of  high-temperature gas dynamics and experiment data have not been fully addressed. It is believed that the rarefaction effects are important in SWBLI, but the systematic analysis of the temperature-jump boundary conditions and the role of translational/rotational/vibrational heat conductivities are lacking. 
In this paper, we derive the three-temperature Navier-Stokes-Fourier (NSF) equations from the gas kinetic theory, with special attention paid to the components of heat conductivity. With proper temperature-jump boundary conditions, we simulate the SWBLI in the double cone experiment.
Our numerical results show that, when the three heat conductivities are properly recovered, the NSF equations can capture the position and peak value of the surface heat flux, in both low- and high-enthalpy inflow conditions. Moreover, the separation bubble induced by the separated shock and the reattachment point induced by impact between transmitted shock and boundary layer are found to agree with the experimental measurement. 
\end{abstract}

\section{Introduction}
\label{sec:introduction}

Aerospace vehicles flying at hypersonic speeds have attracted great attention, where the accurate predictions of heat flux and pressure load on vehicles are one of the key research topics.
The shock wave and boundary layer interaction (SWBLI) has a strong influence upon aerothermodynamic loads on hypersonic vehicles, which has been widely studied experimentally and numerically \citep{GREEN1970, GAITONDE2015, wen2021unsteady,wenzhiyongdc}.
Also, the high-temperature real gas effect is important for aerothermodynamic loads prediction~\citep{AndersonHypersonic} . Usually, the two factors are strongly entangled, and some basic physics are overlooked, resulting in long-lasting discrepancies between the numerical results of high-temperature gas dynamics~\citep{Candler2001} and experimental data~\citep{Knight2018}, e.g., the position and peak value of the heat flux, which are essential in the design of thermal protection system of hypersonic vehicle.


\begin{figure}
	\centering
\includegraphics[viewport=20 4 540 500,clip=true,width=0.45\textwidth]{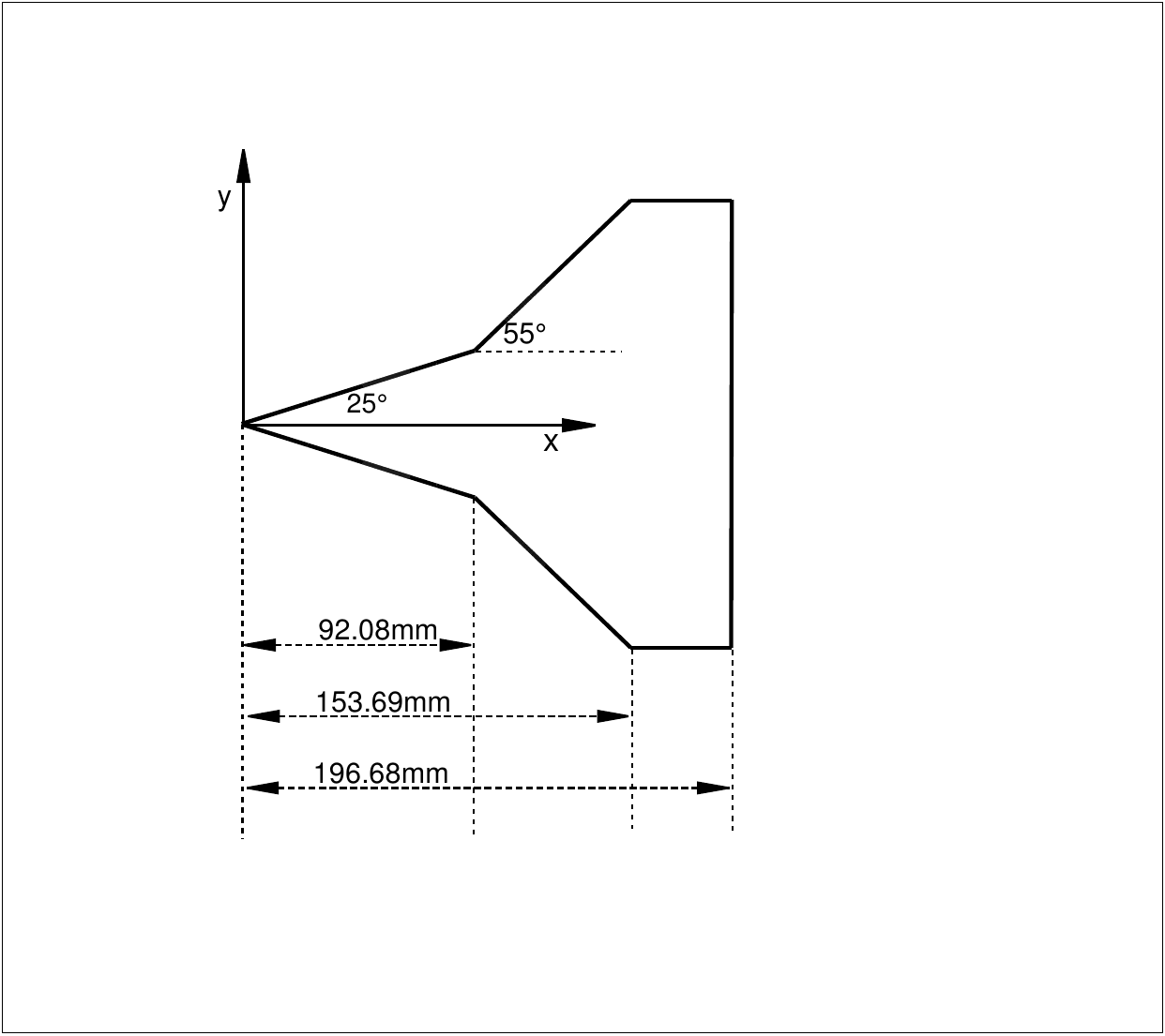}
   \includegraphics[viewport=20 10 540 500,clip=true,width=0.5\textwidth]{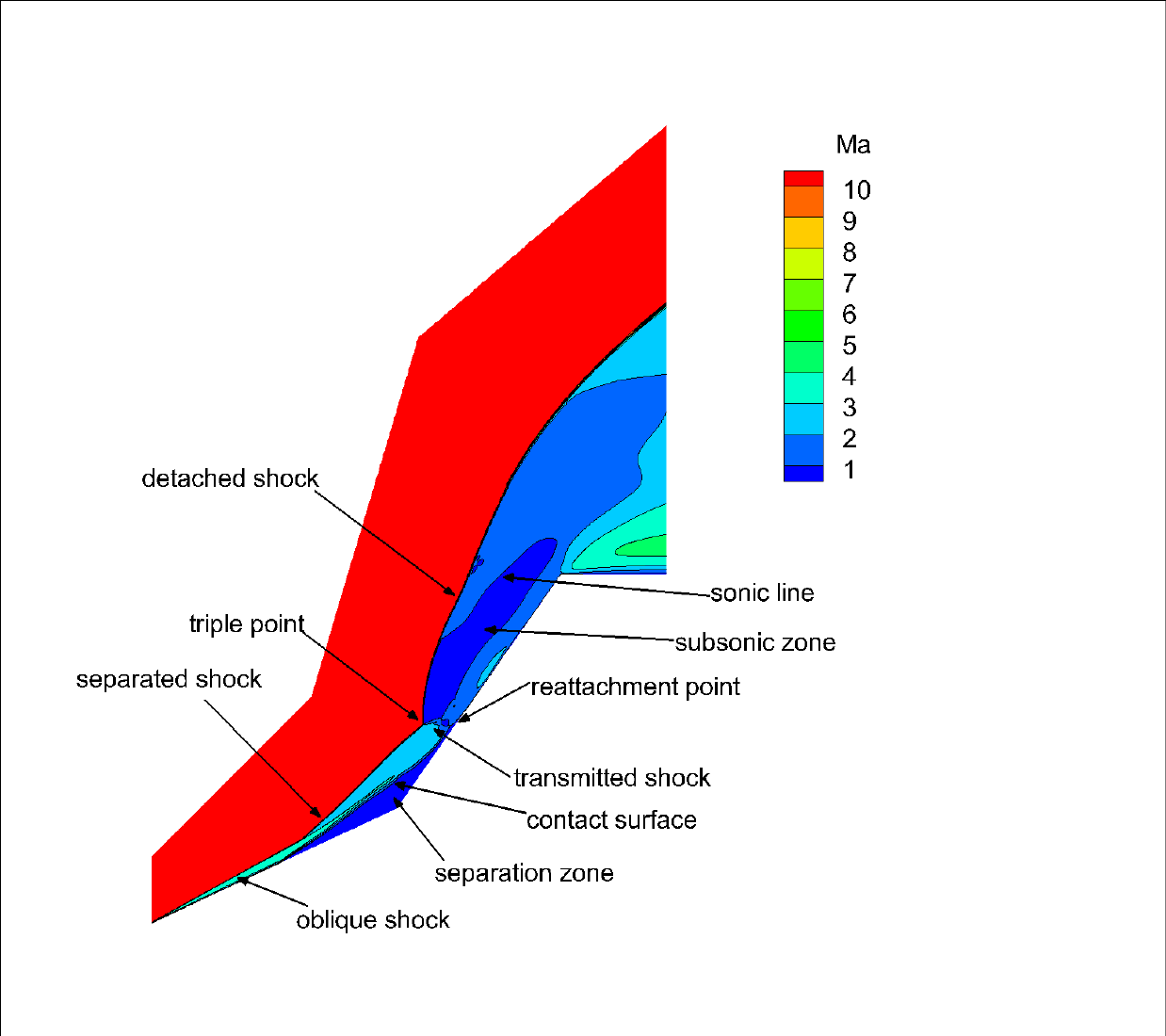}
	\caption{ (Left) Schematic of the $25\degree/55\degree$ axisymmetric double cone. The first and second cone angles are $25\degree$ and $55\degree$, respectively. Coordinate origin locates at the tip of the double cone. (Right) Contour of the Mach number in the Run28 case. A horizontal high-speed inflow comes from the left, generating an attached oblique shock after impinging the first cone, which interacts with a detached bow shock generated by the second cone. A triple point connects separated shock, detached shock and transmitted shock. The impingement between transmitted shock and boundary layer finally results into reattachment which leads to the peak of aerothermodynamic loads. }
	\label{fig:dc_config}
\end{figure}

The double cone configuration is a  benchmarking case for the study of SWBLI~\citep{GAITONDE2015}, e.g., see figure~\ref{fig:dc_config} for  a typical Type-V shock-shock interaction~\citep{EdneyBtype}.
The position and peak value of the surface heat flux, the starting position of the separation bubble,  and the pressure load, are critical factors to assess the accuracy of numerical simulations~\citep{Knight2018}. If the inflow temperature is low, there is no significant high-temperature effects, and the numerical simulations of Navier-Stokes-Fourier (NSF) equations showed good agreement with experiment data \citep{Harvey2001Codevalidation}.   
When the inflow temperature is high, the high-temperature real gas effects are significant, and the NSF equations are difficult to make accurate predictions. 
For instances, for the $25\degree/65\degree$ double cone, \citet{Olejniczak1997Doublecone} found that the simulated separation length is $27\%$ smaller than experiment. 
For the $25\degree/55\degree$ double cone, the computed peak heat flux  is significantly below the experiment data in the Run24 case~\citep{Harvey2001Codevalidation}.
For the Run28 case (see the inflow conditions in table~\ref{tab:fivecasecondition}), \citet{Candler2001} attributed the discrepancy in heat transfer to the differences in characterizing the inflow conditions, such as  the vibrational freezing and the nozzle's non-uniformity.  
For the Run83 case where the stagnation enthalpy is $5.12~ \mathrm{MJ} / \mathrm{kg}$, 
\citet{Nompelis2010} reported the obvious discrepancy in the peak position of heat flux between the numerical and experimental results.
Based on the NSF equations for perfect gas, \citet{Gaitonde2002} did a numerical simulation in nitrogen, and found  that the heat transfer before the separation exceeds the experiment, while the reattachment point is further downstream with experiment.
\citet{Nompelis2005} investigated the non-equilibrium effects in the nozzle numerically. Three vibrational energy relaxation models including Landau-teller relaxation, single quantum relaxation and multi-quantum relaxation are examined for high stagnation enthalpy cases. They found that the peak heat flux is below the experiment data for the Run42 and Run4246 cases. Especially for the Run46c case, the numerical simulation predicts a further down stream location of the separation compared to the experiment data.

\begin{table}
	\begin{center}
		\def~{\hphantom{0}}
		\begin{tabular}{lcccccc}
			Run & Ma & $ H_{t_{\infty}}(\mathrm{MJ} / \mathrm{kg})$ &  $U_{\infty}(\mathrm{m} / \mathrm{s})$ & $T_{\infty}(\mathrm{K})$ & $T_{v ,\infty}(\mathrm{K})$ & $T_w(\mathrm{K})$  \\
			28  & 10.5  &3.4& 2538  & 140.0  & 2589&296.11\\
			35  & 12.59 &3.4& 2545  & 98.27& 2562&296.11\\
			42  & 11.20 &9.1& 4106  & 321.1& 3170&294.72\\
			46  & 11.10 &9.6& 4229  & 347.1& 3083&294.72\\   
		\end{tabular}
		\caption{Inflow conditions \citep{Knight2018} that will be revisited in our numerical simulation.}
		\label{tab:fivecasecondition}
	\end{center}
\end{table}

So far, it is still a challenge to predict the SWBLI with high-temperature real gas effect. On one hand, some scholars believed that the rarefaction effects (e.g. the temperature-jump boundary condition) are important in the SWBLI carried out in the wind tunnel experiment. On the other hand, inspired by our recently work in the rarefied gas dynamics of polyatomic gas~\citep{li2021uncertainty}, we conjecture that the inaccurate use of the translational/rotational/vibrational heat conductivities is also a major reason to the discrepancy in numerical and experimental results. However, the systematic analysis is lacking.
In this paper, 
the experimental data of hypersonic flows past double cone~\citep{Knight2018} are used to reveal the role of thermal conductivities and boundary condition. As pointed out by~\cite{Kustova2022}, we only consider the
thermodynamic non-equilibrium effects rather than the chemical reactions  because the temperature is not high enough in wind tunnel experiment.

\section{Mathematical formulation for the governing equation}\label{sec:equation}

While the total heat conductivity can be relatively easily measured from experiment, it is a difficult task to distinguish the relative proportion of translational, rotational, and vibrational heat conductivities.  
As a consequence, the current method based on NSF equations can predict the low-temperature SWBLI well~\citep{Harvey2001Codevalidation}, but fails to predict the high-temperature gas dynamics. Therefore, it is of great importance to drive the multi-temperature NSF equations from the gas kinetic theory. 

Physically, the non-equilibrium dynamics of dilute polyatomic gas is described by the \cite{wangcs1951transport} (WCU) equation, an extended version of the Boltzmann equation for polyatomic gases; however, it is intractable in practical problems due to its computational complexity and the lack of information of transition probabilities (in the Boltzmann collision kernel) between energy levels. To bypass this difficulty, various gas kinetic model equations have been proposed by imitating the properties of the WCU equation with simplified collision terms \citep{Morse1964,rykov1975model,andries2000gaussian,LeiJFM2015}. Nevertheless, when the Knudsen number $\text{Kn}$ (i.e., the ratio between the mean free path of gas molecules and characteristic flow length) is small, the dynamics of polyatomic gas can be well described by the multi-temperature macroscopic equations, together with the velocity-slip and temperature-jump boundary conditions \citep{Aoki_NSF_BC0,Aoki_NSF_BC,Su2022PoF}, and thus the computational complexity can be reduced significantly. Such formal derivation of macroscopic equations can be facilitated by the Chapman-Enskog expansion method \citep{CE} based on either Boltzmann/WCU equations or kinetic model equations. However, the former leads to the transport properties in the NSF equations involving complicated collision integrals \citep{Kustova2019,Kustova2020}, and the latter is only limited to polyatomic gas with single internal energy mode based on the ellipsoidal-statistical model \citep{Aoki_NSF2020}, where the heat conductivity cannot be adjusted once the bulk viscosity is determined. Fortunately, the modified Rykov model was recently constructed for polyatomic gases with rotational and vibrational energies, where all the transport properties, particularly the heat conductivity of each internal mode, can be correctly and independently recovered \citep{JFM2023Li}. Therefore, we will derive a set of three-temperatures NSF equations from the modified Rykov model to serve as the governing equations in the present work.

\subsection{Kinetic model equation}

We consider the polyatomic gas with rotational and vibrational degrees of freedom  $d_r$ and $d_v(T_v)$, respectively, where $d_r$ is a constant since the rotational degrees of freedom is fully excited in experimental conditions, while $d_v$ changes with the vibrational temperature $T_v$ as,
\begin{equation}\label{eq:harmonic_oscillator_dv}
	\begin{aligned}[b]
		d_v(T_v)=\frac{2\theta_v/T_v}{\exp({\theta_v/T_v})-1},
	\end{aligned}
\end{equation}
with $\theta_v$ being the characteristic temperature of the active vibrational mode.

The distribution function $f(t,\bm{x},\bm{v},I_r,I_v)$ is used to identify the states of polyatomic gas, where $t$ is the time, $\bm{x}$ is the spatial coordinates, $\bm{v}$ is the molecular velocity, $I_r$ and $I_v$ are the rotational and vibrational energy, respectively. Then, the macroscopic variables, such as the mass density $\rho$, flow velocity $\bm{u}$, heat fluxes $\bm{q}_t,\bm{q}_r,\bm{q}_v$, pressure tensor $P$ and temperatures $T_t,T_r,T_v$, are obtained by taking the moments of the distribution function:
\begin{equation}\label{eq:macroscopic_variables_f}
	\begin{aligned}[b]
		\left(\rho, \rho\bm{u},P\right)=\int\left(1,\bm{v},\bm{c}\bm{c}\right){f}\mathrm{d}\bm{v}\mathrm{d}I_r\mathrm{d}I_v, \\
		\left(\frac{3}{2}\rho RT_t,\frac{d_r}{2}\rho RT_r,\frac{{d_v(T_v)}}{2}\rho RT_v\right)=\int{\left(\frac{1}{2}c^2,I_r,I_v\right)f}\mathrm{d}\bm{v}\mathrm{d}I_r\mathrm{d}I_v, \\
		\left(\bm{q}_t,\bm{q}_r,\bm{q}_v\right)=\int\bm{c}\left(\frac{1}{2}c^2,I_r,I_v\right){f}\mathrm{d}\bm{v}\mathrm{d}I_r\mathrm{d}I_v,
	\end{aligned}
\end{equation}
where the subscripts $t,~r,~v$ indicate the translational, rotational and vibrational component, respectively; $\bm{c}=\bm{v}-\bm{u}$ is the thermal velocity, $R$ is the specific gas constant, and the integration range for the velocity, rotational energy, vibration energy are $(-\infty,\infty)^3$, $(0,\infty)$, $(0,\infty)$, respectively. The temperatures $T_{tr}$, $T_{tv}$ and $T$ are defined as the equilibrium temperature between translational and rotational modes, translational and vibrational modes, and overall modes, respectively,
\begin{equation}\label{eq:Ttr_Ttv_T}
	\begin{aligned}[b]
		T_{tr}=\frac{3T_t+d_rT_r}{3+d_r}, \\ T_{tv}=\frac{3T_t+{{d_v(T_v)}}T_v}{3+{{d_v(T_{tv})}}}, \\ T=\frac{3T_t+d_rT_r+{{d_v(T_v)}}T_v}{3+d_r+{{d_v(T)}}},
	\end{aligned}
\end{equation}
and the corresponding pressures are $(p_t, p_r, p_v, p, p_{tr}, p_{tv}) = \rho R(T_t, T_r, T_v, T, T_{tr}, T_{tv})$.

When the system is in equilibrium, we have 
the equilibrium distribution function $f^{eq}=f^{eq}_t(T)f^{eq}_r(T)f^{eq}_v(T)$, where 
\begin{equation}\label{eq:Et_Er_Ev}
\begin{aligned}[b]
f^{eq}_t(T)&=\rho{\left(\frac{1}{2\pi RT}\right)}^{3/2}\exp{\left(-\frac{c^2}{2RT}\right)}, \\
f^{eq}_r(T)&=\frac{I^{d_r/2-1}_{r}}{\Gamma(d_r/2)(RT)^{d_r/2}}\exp{\left(-\frac{I_r}{RT}\right)}, \\
f^{eq}_v(T)&=\frac{I^{{{d_v(T)}}/2-1}_{v}}{\Gamma({{d_v(T)}}/2)(RT)^{d_v(T)/2}}\exp{\left(-\frac{I_v}{RT}\right)}.
\end{aligned}
\end{equation}
and $\Gamma$ is the gamma function. 

But when the system is out-of-equilibrium,  the evolution of the distribution function (in the absence of external force) is governed by the following kinetic equations:
\begin{equation}\label{eq:kinetic_model_equation}
  \frac{\partial{f}}{\partial{t}}+\bm{v} \cdot \frac{\partial{f}}{\partial{\bm{x}}}
  = \frac{g_t-f}{\tau} + \frac{g_r-g_t}{Z_r\tau} + \frac{g_v-g_t}{Z_v\tau},
\end{equation}
where $\tau$ is the characteristic time of elastic collisions, $Z_r$ and $Z_v$ are the rotational and vibrational collision number (their products with $\tau$ characterizing the rotational and vibrational relaxation times), respectively. The reference distribution functions $g_t,~g_r,~g_v$ are expanded about the equilibrium distribution $f^{eq}$ in a series of orthogonal polynomials in variables $\bm{c}$, $I_r$, $I_v$ and $\bm{q_t},~\bm{q_r},~\bm{q_v}$:
\begin{align}\label{eq:gt_gr_gv}
	g_t =~&f^{eq}_t(T_t) f^{eq}_r(T_r) f^{eq}_v(T_v) \left[{1 + \frac{2\bm{q}_t\cdot{\bm{c}}}{15{R}{T_t}{p_t}} \left(\frac{c^2}{2RT_t}-\frac{5}{2}\right)} \right. \notag \\ 
	&\left. {+\frac{2\bm{q}_r\cdot{\bm{c}}}{d_rRT_tp_r}\left(\frac{I_r}{RT_r}-\frac{d_r}{2}\right) + \frac{2\bm{q}_v\cdot{\bm{c}}}{{{d_v(T_v)}}RT_tp_v}\left(\frac{I_v}{RT_v}-\frac{{d_v(T_v)}}{2}\right)}\right], \notag \\
	g_r =~&f^{eq}_t(T_{tr}) f^{eq}_r(T_{tr}) f^{eq}_v(T_v) \left[{1 + \frac{2\bm{q}_0\cdot{\bm{c}}}{15{R}{T_{tr}}{p_{tr}}} \left(\frac{c^2}{2RT_{tr}}-\frac{5}{2}\right)} \right. \notag \\ 
	&\left. {+\frac{2\bm{q}_1\cdot{\bm{c}}}{d_rRT_{tr}p_{tr}}\left(\frac{I_r}{RT_{tr}}-\frac{d_r}{2}\right) + \frac{2\bm{q}_2\cdot{\bm{c}}}{{{d_v(T_v)}}RT_{tr}p_v}\left(\frac{I_v}{RT_v}-\frac{{d_v(T_v)}}{2}\right)}\right], \notag \\
	g_v =~&f^{eq}_t(T_{tv}) f^{eq}_r(T_{r}) f^{eq}_v(T_{tv}) \left[{1 + \frac{2\bm{q}_0\cdot{\bm{c}}}{15{R}{T_{tv}}{p_{tv}}} \left(\frac{c^2}{2RT_{tv}}-\frac{5}{2}\right)} \right. \notag \\ 
	&\left. {+\frac{2\bm{q}_1\cdot{\bm{c}}}{d_rRT_{tv}p_{r}}\left(\frac{I_r}{RT_{r}}-\frac{d_r}{2}\right) + \frac{2\bm{q}_2\cdot{\bm{c}}}{{{d_v(T_{tv})}}RT_{tv}p_{tv}}\left(\frac{I_v}{RT_{tv}}-\frac{{d_v(T_{tv})}}{2}\right)}\right], 
\end{align}
where ${\bm{q}_{0}}$, ${\bm{q}_{1}}$, and ${\bm{q}_{2}}$ are linear combinations of translational, rotational and vibrational heat fluxes:
\begin{equation}\label{eq:q0_q1_q2}
  \begin{bmatrix} 
    \bm{q}_{0} \\ \bm{q}_{1} \\ \bm{q}_{2}
\end{bmatrix}
= 
  \begin{bmatrix}		(2-3A_{tt})Z_{int}+1 & -3A_{tr}Z_{int} & -3A_{tv}Z_{int} \\		-A_{rt}Z_{int} & -A_{rr}Z_{int}+1 & -A_{rv}Z_{int} \\ 		-A_{vt}Z_{int} & -A_{vr}Z_{int} & -A_{vv}Z_{int}+1
  \end{bmatrix}
  \begin{bmatrix} 
    \bm{q}_{t} \\ \bm{q}_{r} \\ \bm{q}_{v}
  \end{bmatrix},
\end{equation}
with $Z_{int}=\left({1}/{Z_r}+{1}/{Z_v}\right)^{-1}$, and the matrix $\bm{A}=A_{ij}~(i,j=t,r,v)$ determined by the relaxation rates of heat flux~\citep{li2021uncertainty,JFM2023Li}. 

\subsection{Three-temperature macroscopic equations}

When $Z_r\sim Z_v\sim O(1)$, the internal energy exchange with the translational energy is quick, so the standard Chapman-Enskog expansion method of the gas kinetic equation leads to one-temperature NSF equations; compared to the traditional NSF equations without bulk viscsoity,  for polyatomic gas the bulk viscosity appearing in the constitutive relation of stress \citep{CE}. However, the energy exchange between the translational and rotational/vibrational modes usually takes much longer time than the elastic collisions, particularly for translational-vibrational relaxation, which renders the one-temperature NSF equations invalid. Therefore, for practical consideration, we assume that $Z_r\sim O(\text{Kn}^{-1}), Z_v\sim O(\text{Kn}^{-2})$, and apply the Chapman-Enskog method to find the three-temperature NSF equations and the associated transport coefficients of the slightly-rarefied polyatomic gas up to the slip-flow limit, that is, $\text{Kn} \lessapprox 0.01$.

Taking the moments of the kinetic equation \eqref{eq:kinetic_model_equation}, we obtain 
\begin{equation}\label{eq:macroscopic_equation}
	\begin{aligned}[b]
		\frac{\partial{\rho}}{\partial{t}} + \nabla\cdot\left(\rho\bm{u}\right) &= 0, \\
		\frac{\partial}{\partial{t}}\left(\rho\bm{u}\right) + \nabla\cdot\left(\rho\bm{u}\bm{u}\right) + \nabla\cdot\bm{P} &= 0, \\
    \frac{\partial}{\partial{t}}\left(\rho e\right) + \nabla\cdot\left(\rho e\bm{u}\right) + \nabla\cdot\left(\bm{P}\cdot\bm{u}+\bm{q}_t+\bm{q}_r+\bm{q}_v\right) &= 0, \\
    \frac{\partial}{\partial{t}}\left(\rho e_r\right) + \nabla\cdot\left(\rho e_r\bm{u}\right) + \nabla\cdot\bm{q}_r &= \frac{\rho R}{2}\frac{d_r\left(T_{tr}-T_r\right)}{Z_r\tau}, \\
    \frac{\partial}{\partial{t}}\left(\rho e_v\right) + \nabla\cdot\left(\rho e_v\bm{u}\right) + \nabla\cdot\bm{q}_v &= \frac{\rho R}{2}\frac{d_v(T_{tv})T_{tv}-d_v(T_{v})_v}{Z_v\tau}, \\
	\end{aligned}
\end{equation} 
where $e_r=d_rRT_r/2$, $e_v=d_v(T_v)RT_v/2$ and $e=(3RT_t+u^2)/2+e_r+e_v$ are the specific rotational, vibrational and total energies, respectively. To close the above set of equations, it is necessary to find the approximations to distribution function $f$, and thus the pressure tensor $\bm{P}$ and heat fluxes $\bm{q}_t,\bm{q}_r,\bm{q}_v$ can be expressed as functions of the macroscopic variables $\rho,\bm{u},T_t,T_r,T_v$.

In the Chapman-Enskog method, the distribution function $f$ is an expansion in the form of an infinite series of $\epsilon$ in the continuum limit ($\epsilon$ has the same order of $\text{Kn}$),
\begin{equation}\label{eq:f_expansion}
	\begin{aligned}[b]
		f = f^{(0)} + \epsilon f^{(1)} + \epsilon^2 f^{(2)} + \cdots .\\
	\end{aligned}
\end{equation} 
The unexpanded macroscopic variables $\rho,\bm{u},T_t,T_r,T_v$ are determined only by $f^{(0)}$, while the other macroscopic quantities $(h=\bm{P},\bm{q}_t,\bm{q}_r,\bm{q}_v)$ are also expanded as,
\begin{equation}\label{eq:h_expansion}
	\begin{aligned}[b]
		h = h^{(0)} + \epsilon h^{(1)} + \epsilon^2 h^{(2)} + \cdots .\\
	\end{aligned}
\end{equation} 
Substituting the expansions into the model equation \eqref{eq:kinetic_model_equation} with the assumption $Z_r\sim O(\epsilon^{-1})$ and $Z_v\sim O(\epsilon^{-2})$, the first approximation (zero-order) to distribution functions $f^{(0)}$ is given by the solution of the kinetic equation
\begin{equation}\label{eq:0th_equation}
	\begin{aligned}[b]
		\mathcal{D}^{(0)}f = \frac{g_t-f^{(0)}}{\tau},
	\end{aligned}
\end{equation} 
where $\mathcal{D}f\equiv{\partial{f}}/{\partial{t}}+\bm{v} \cdot {\partial{f}}/{\partial{\bm{x}}}$, and $\mathcal{D}^{(0)}f=0$. The vanishing of the collision term indicates that $f^{(0)}$ is a local equilibrium distribution function with the temperatures $T_t,T_r,T_v$ of respective modes,
\begin{equation}\label{eq:f0}
	\begin{aligned}[b]
		f^{(0)} = f^{eq}_t(T_t) f^{eq}_r(T_r) f^{eq}_v(T_v).
	\end{aligned}
\end{equation} 
Then the zero-order pressure $\bm{P}^{(0)}$ and heat fluxes $\bm{q}^{(0)}_t,\bm{q}^{(0)}_r,\bm{q}^{(0)}_v$ can be obtained by taking moments of $f^{(0)}$,
\begin{equation}\label{eq:P_q_0th}
	\begin{aligned}[b]
		\bm{P}^{(0)} &= \int{\bm{c}\bm{c}f^{(0)}}\mathrm{d}\bm{v}\mathrm{d}I_r\mathrm{d}I_v = \rho RT_t\bm{\mathrm{I}}, \\
    \left(\bm{q}^{(0)}_t,\bm{q}^{(0)}_r,\bm{q}^{(0)}_v\right) &=\int\bm{c}\left(\frac{1}{2}c^2,I_r,I_v\right)f^{(0)}\mathrm{d}\bm{v}\mathrm{d}I_r\mathrm{d}I_v = \left(0,0,0\right),
	\end{aligned}
\end{equation} 
where $\bm{\mathrm{I}}$ is the identity matrix. The zero-order pressure and heat fluxes give a set of governing equations in the Euler approximation. 

To the second approximation of the distribution function $f^{(0)}+\epsilon f^{(1)}$, the first-order correction $f^{(1)}$ is solved from the kinetic equations:
\begin{equation}\label{eq:1st_equation}
	\begin{aligned}[b]
		\mathcal{D}^{(1)}f = -\frac{f^{(1)}}{\tau} + \frac{g_r-f^{(0)}}{Z_r\tau},
	\end{aligned}
\end{equation}
where $\mathcal{D}^{(1)}f={\partial{f^{(0)}}}/{\partial{t}}+\bm{v} \cdot {\partial{f^{(0)}}}/{\partial{\bm{x}}}$  can be explicitly evaluated:
\begin{equation}\label{eq:D1f}
	\begin{aligned}[b]
		\mathcal{D}^{(1)}f=f^{(0)}&\left[\left(\frac{c^{2}}{2RT_{t}}-\frac{5}{2}\right)\bm{c}\cdot\nabla\ln{T_{t}} +\left(\frac{I_r}{RT_{r}}-\frac{d_r}{2}\right)\bm{c}\cdot\nabla\ln{T_{r}} +\left(\frac{I_v}{RT_v} -\frac{d_v}{2} \right)\bm{c}\cdot\nabla\ln{T_v} \right. \\
		&\left. +\left(\frac{d_r}{3T_{t}}\left(\frac{c^{2}}{2RT_{t}}-\frac{3}{2}\right) -\frac{1}{T_r}\left(\frac{I_r}{RT_r}-\frac{d_r}{2}\right)\right)\frac{T_{tr}-T_r}{Z_r\tau} \right. \\
		&\left. +\frac{1}{RT_{t}}\left(\bm{c}\bm{c}-\frac{1}{3}c^2\mathrm{I}\right):\nabla\bm{u} \right] + O(\epsilon^{2}).
	\end{aligned}
\end{equation}
Therefore, the first-order correction of the distribution function $f^{(1)}$ is obtained. Substituting the second approximation $f^{(0)}+\epsilon f^{(1)}$ into the definitions of the pressure tensor and heat fluxes, the constitutive relations at the NSF level are obtained as follows:
\begin{equation}\label{eq:P_q_NSF}
	\begin{aligned}[b]
		\bm{P}^{\text{NSF}} &= \rho RT_t\bm{\mathrm{I}} -\mu\left(\nabla\bm{u}+\nabla\bm{u}^{\mathrm{T}}-\frac{2}{3}\nabla\cdot\bm{u}\bm{\mathrm{I}}\right), \\
    \left(\bm{q}^{\text{NSF}}_t,\bm{q}^{\text{NSF}}_r,\bm{q}^{\text{NSF}}_v\right) &= -\left(\kappa_t\nabla{T_{t}},\kappa_r\nabla{T_{r}},\kappa_v\nabla{T_{v}}\right),
	\end{aligned}
\end{equation} 
where the shear viscosity is
\begin{equation}
\mu = \rho RT_t\tau,
\end{equation} 
and heat conductivities $\kappa_t,\kappa_r,\kappa_v$ are given by
\begin{equation}\label{eq:mu_kappa}
	\begin{aligned}[b]
    \left[ 
      \begin{array}{ccc} 
        \kappa_t \\ \kappa_r \\ \kappa_v
      \end{array}
    \right]
	&= \frac{\mu R}{2}
    \left[ 
      \begin{array}{ccc} 
        A_{tt} & A_{tr} & A_{tv} \\ A_{rt} & A_{rr} & A_{rv} \\ A_{vt} & A_{vr} & A_{vv}
      \end{array}
    \right]^{-1}
    \left[ 
      \begin{array}{ccc} 
        5 \\ d_r \\ d_v(T_v)
      \end{array}
    \right].
	\end{aligned}
\end{equation} 

It should be noted that (i) the bulk viscosity does not appear in \eqref{eq:P_q_NSF}. This is because for dilute gas the bulk viscosity originates from the rotational/vibrational-translational energy exchange, and this is already included in the last two equations in \eqref{eq:macroscopic_equation}; If $Z_r$ and $Z_v$ is at the order of $O(1)$, then the difference between the rotational/vibrational temperatures and translational temperature is small, and one can perform the Chapman-Enskog expansion once to the equation \eqref{eq:P_q_NSF}, to remove the last two equations but to add the bulk viscosity to the stress.  (ii)  from \eqref{eq:P_q_NSF} we know that,  each component of the heat flux is related to the corresponding temperature gradient of its own mode, due to the thermal non-equilibrium when translational-internal energy relaxation is slow. It should be noted that the heat conductivities given by the modified Rykov model are determined by the thermal relaxation rates of heat flux, and hence can be adjusted independently to model the realistic heat transport properties of each mode of a polyatomic gas.

\subsection{Determination of parameters}

The three-temperature NSF equations, given by \eqref{eq:macroscopic_equation} and \eqref{eq:P_q_NSF}-\eqref{eq:mu_kappa}, have been established, which contain the collision numbers $Z_r,Z_v$, the shear viscosity $\mu$ and the heat conductivities $\kappa_t,\kappa_r,\kappa_v$. Although the transport parameters can be calculated straightforwardly based on the gas kinetic theory, it is prevented by the complexity and computational burden, while semi-empirical values are commonly adopted for practical flow problems. 

The viscosities of the components of air at high temperatures have been given by \cite{Gupta11speciedairreaction}, and the expression for nitrogen is,
\begin{equation}\label{miucurvefitting}
  \mu(T_t)=\exp (-11.8153) T_t^{0.0203 \ln T_t+0.4329}.
\end{equation}
For the collision numbers, \cite{Parker1959} presented an analytical derivation of $Z_r$, which shows good predictions of shock wave structures with the parameters \citep{Lordi1970PoF},
\begin{equation}
	\begin{aligned}[b]
		Z_{r}=\frac{Z_{r}^{\infty}}{1+\frac{\sqrt{\pi}}{2} \sqrt{\frac{T^{*}}{T_{t}}}+\left(\pi+\frac{\pi^{2}}{4}\right) \frac{T^{*}}{T_{t}}},\quad Z_{r}^{\infty}=23.0, 
		\quad
		T^{*}=91.5 \mathrm{~K}.
	\end{aligned}
\end{equation}
Also, the most commonly used empirical model for temperature-dependent vibrational collision number $Z_v$ for diatomic molecules is given by \cite{Millikan1963jcp}:
\begin{equation}\label{vibrational_collision_number}
	\begin{aligned}[b]
		Z_v=\frac{C_1}{T_v^{\omega}}\exp{\left(\frac{C_2}{T_v^{1/3}}\right)},
		\quad C_{1}=9.1 ,
		\quad C_{2}=220.0,
	\end{aligned}
\end{equation}
where $\omega$ is the viscosity index and the temperature is given in the units of Kelvin. 

On the other hand, the accurate determination of heat conductivity components and their influence in predicting heat transfer is always overlooked~\citep{Wu2020JFM,li2021uncertainty}. In the thermal equilibrium system, giving total heat conductivity without the details on the proportions of different modes is adequate to determine the heat transfer. However, as shown in \eqref{eq:P_q_NSF}, the heat flux depends on the temperature gradient of each energy mode. Therefore, even when the total heat conductivity is fixed, different proportion of the translational, rotational and vibrational heat conductivities lead to different total heat flux. Therefore, the translational, rotational and vibrational heat conductivities have to be correctly provided. 

Historically, the dimensionless \cite{EUCKEN1913} factor $f_{eu}$ is used to quantify the ratio of heat conductivity to shear viscosity of a gas,
\begin{equation}\label{eq:feu}
	\begin{aligned}[b]
		c_vf_{eu}\equiv\frac{\kappa}{\mu}=\frac{\kappa_t+\kappa_r+\kappa_v}{\mu},
	\end{aligned}
\end{equation}
where $\kappa$ is the total heat conductivity,  and $c_v$ is the total specific heat capacity at constant volume. Similarly, $f_t$, $f_r$, and $f_v$ represent the Eucken factors of the translational, rotational, and vibrational modes, respectively,
\begin{equation}\label{eq:ft_fr_fv}
	\begin{aligned}[b]
		f_t = \frac{\kappa_t}{c_{v,t}\mu}, \quad f_r = \frac{\kappa_r}{c_{v,r}\mu}, \quad f_v = \frac{\kappa_v}{c_{v,v}\mu},
	\end{aligned}
\end{equation}
where $c_{v,t},c_{v,r},c_{v,v}$ are the specific heat capacities at constant volume for each mode. Therefore, based on the constitutive relation of heat flux \eqref{eq:mu_kappa}, the Eucken factors are determined by the thermal relaxation rates $\bm{A}$ as
\begin{equation}\label{eq:EuckenFactor_A}
\left[ 
\begin{array}{ccc} 
f_t \\ f_r \\ f_v
\end{array}
\right]
= 
\left[ 
\begin{array}{ccc} 
3A_{tt} & d_rA_{tr} & d_v(T_v)A_{tv} \\ 3A_{rt} & d_rA_{rr} & d_v(T_v)A_{rv} \\ 3A_{vt} & d_rA_{vr} & d_v(T_v)A_{vv}
\end{array}
\right]^{-1}
\left[ 
\begin{array}{ccc} 
5 \\ d_r \\ d_v(T_v)
\end{array}
\right].
\end{equation}

Despite that the total Eucken factor $f_{eu}$ can be measured experimentally, the determination of its components is rather difficult, due to the difficulty in obtaining the relaxation matrix $A$~\citep{JFM2023Li}. Initially, by analogy with noble gases and simple gas kinetic theory, \cite{EUCKEN1913} assigned the translational and internal Eucken factors as 
$f_{t}=\frac{5}{2}$ and  $f_{r}=f_{v}=1$.
Later, it was realized that the transport of internal energy occurs by the molecular diffusion mechanism \citep{Hirschfelder1957}, thus $f_{r}$ and $f_{v}$ are expressed in terms of the self-diffusion coefficient $D$,
\begin{equation}\label{Euken_2}
	\begin{aligned}[b]
		f_{t}=\frac{5}{2}, \quad f_{r}=f_{v}=\frac{\rho D}{\mu}\equiv \frac{1}{Sc},
	\end{aligned}
\end{equation}
where the Schmidt number $Sc$ of nitrogen is around $1/1.33$ in a wide range of temperatures. Finally, by using the asymptotic expansion of the WCU equation, \cite{MasonandMonchick1962Eucken} derived more accurate expressions for the translational and internal Eucken factors. Based on their results, the Eucken factors are in general functions of temperature, and the curve-fitting formulations of Eucken factors are calculated for nitrogen as:
\begin{equation}\label{ft_fint_fv}
  \begin{aligned}[b]
    f_{t}&=\left\{
      \begin{array}{cc}
        1.80  , T_{t} \leq 100 \mathrm{~K}, \\
        \frac{2.435 T_{t}^{3}+47.1 T_{t}^{2}+3.42 T_{t}+0.8337}{T_{t}^{3}+61.13 T_{t}^{2}-4.624 T_{t}+0.8556}, 100 \mathrm{~K}<T_{t} \leq 3000 \mathrm{~K}, \\
        2.42 , T_{t}>3000 \mathrm{~K}, \\
      \end{array}\right. \\
    f_{int}&=\left\{
      \begin{array}{cc}
        1.88  , T_{int} \leq 100 \mathrm{~K}, \\
        \frac{1.373 T_{int}^{3}+125.7 T_{int}^{2}-3.87 T_{int}+0.2182}{T_{int}^{3}+39.89 T_{int}^{2}+8.668 T_{int}+0.9401}, 100 \mathrm{~K}<T_{int} \leq 3000 \mathrm{~K}, \\
        1.40 , T_{int}>3000 \mathrm{~K},\\
      \end{array}\right. \\
    f_v&=1.37,
  \end{aligned}
\end{equation}
where $T_{int}={\left(d_{r} T_{r}+d_{v} T_{v}\right)}/{\left(d_{r}+d_{v}\right)}$, and then the rotational Eucken factor $f_r$ can be obtained by the relation $f_{r} c_{v,r}+f_{v} c_{v,v}=f_{int}\left(c_{v,r}+c_{v,v}\right)$.

%
%

\subsection{Velocity slip and temperature jump}

Due to the low-density gas used in the wind tunnel, the flow can reach a slip-flow regime. Therefore, in addition to multi-temperature NSF equations, the velocity-slip and temperature-jump boundary conditions at the wall surfaces have to be specified. 

From the rigorous derivations based on kinetic theory and the Maxwell model of gas-wall interaction, both the velocity slip and temperature jump are affected by the velocity and temperature gradients on the surfaces \citep{Aoki_NSF_BC0,Aoki_NSF_BC}. Nevertheless, when an isothermal wall is considered, the velocity slip can be much simplified by a first-order extrapolation of the values beyond the Knudsen layer, 
\begin{equation}\label{velocity_slip}
  u_t-u_{w,t}=\frac{2-\alpha_t}{\alpha_t}\lambda \frac{\partial u_t}{\partial n},
\end{equation}
where $u_t$ and $u_{w,t}$ are the tangential velocities of the gas and surface, respectively, and ${\partial}/{\partial n}$ is the gradient along the unit normal vector $\bm{n}$ pointing toward the flowfield; $\alpha_t$ is the tangential momentum accommodation coefficient, and $\lambda$ is the local mean free path of gas molecules calculated by
\begin{equation}
\lambda=\frac{\mu(T_t)}{\rho} \sqrt{\frac{\pi}{2RT_t}}.
\end{equation}
Meanwhile, the component of the velocity normal to the wall is always zero.

The temperature jump condition is relatively more difficult to  establish for polyatomic gases, since in general it depends on the translational-internal energy exchange rate, as well as the heat conductivity components \citep{Su2022PoF}. Similar to the velocity slip condition, the first-order temperature jump condition of a monatomic gas with a rough estimation of the jump coefficient is given by,
\begin{equation}\label{eq:temperature_jump_T}
  T-T_{w}=\frac{2\gamma}{(\gamma+1)\text{Pr}}\frac{2-\sigma}{\sigma}\lambda \frac{\partial T}{\partial n},
\end{equation}
where $T_w$ is the wall temperature, $\sigma$ is the energy accommodation coefficient, $\gamma$ is the specific heat ratio, and $\text{Pr}$ is the Prandtl number. However, application of  \eqref{eq:temperature_jump_T} to all energy modes predicts a significantly higher vibrational heat flux, and thus \cite{Candler2002} used a modified vibrational energy jump as a simple extension,
\begin{equation}\label{eq:temperature_jump_candler}
  e_v-e_{w,v}=\frac{2-\sigma_{v}}{\sigma_{v}}\lambda \frac{\partial e_v}{\partial n},
\end{equation}
where $\sigma_{v}$ is the energy accommodation coefficient for vibrational mode, which is suggested to be a very small value ($1.0\times10^{-3}$ typically) to model the nearly adiabatic condition of vibrational mode. However, the value of the energy accommodation coefficient of each energy mode is practically difficult to be known.

We noticed that $\frac{2-\sigma_{v}}{\sigma_{v}} \approx 2000$ which is close to $Z_v$. Therefore, to take into account the different degrees of accommodation of translational, rotational and vibrational modes, we propose the following temperature jump boundary conditions for thermal non-equilibrium flows of polyatomic gases:
\begin{equation}\label{eq:temperature_jump_current}
    \begin{aligned}[b]
      T_{t}-T_{w}&=\frac{2-\sigma}{\sigma}\lambda \frac{\partial T_t}{\partial n},\\
      T_{r}-T_{w}&=\frac{2-\sigma}{\sigma}Z_{r} \lambda \frac{\partial T_r}{\partial n}, \\
      T_{v}-T_{w}&=\frac{2-\sigma}{\sigma} Z_{v} \lambda \frac{\partial T_v}{\partial n},
\end{aligned}
\end{equation}
where $\sigma$ is the common energy accommodation coefficient, while the rotational and vibrational collision numbers appearing in the corresponding temperature jump coefficients, to take into account the slower relaxation between internal energies of the gas and the wall. 
In the simulation, $\alpha_t$ and $\sigma$ take the value of 1.

\section{Numerical results}\label{sec:results}

The numerical solutions of the multi-temperature NSF equations with the velocity-slip and temperature-jump boundary conditions will be compared to the experimental data for four cases summarized in table \ref{tab:fivecasecondition}, where the axisymmetric configuration is shown in figure \ref{fig:dc_config}.  
In this section, the Eucken factors  \eqref{eq:ft_fr_fv} are calculated from \eqref{ft_fint_fv}.

\begin{figure}
	\centering 
	\subfigure[translational temperature]{\includegraphics[viewport=20 10 540 500,clip=true,width=0.485\textwidth]{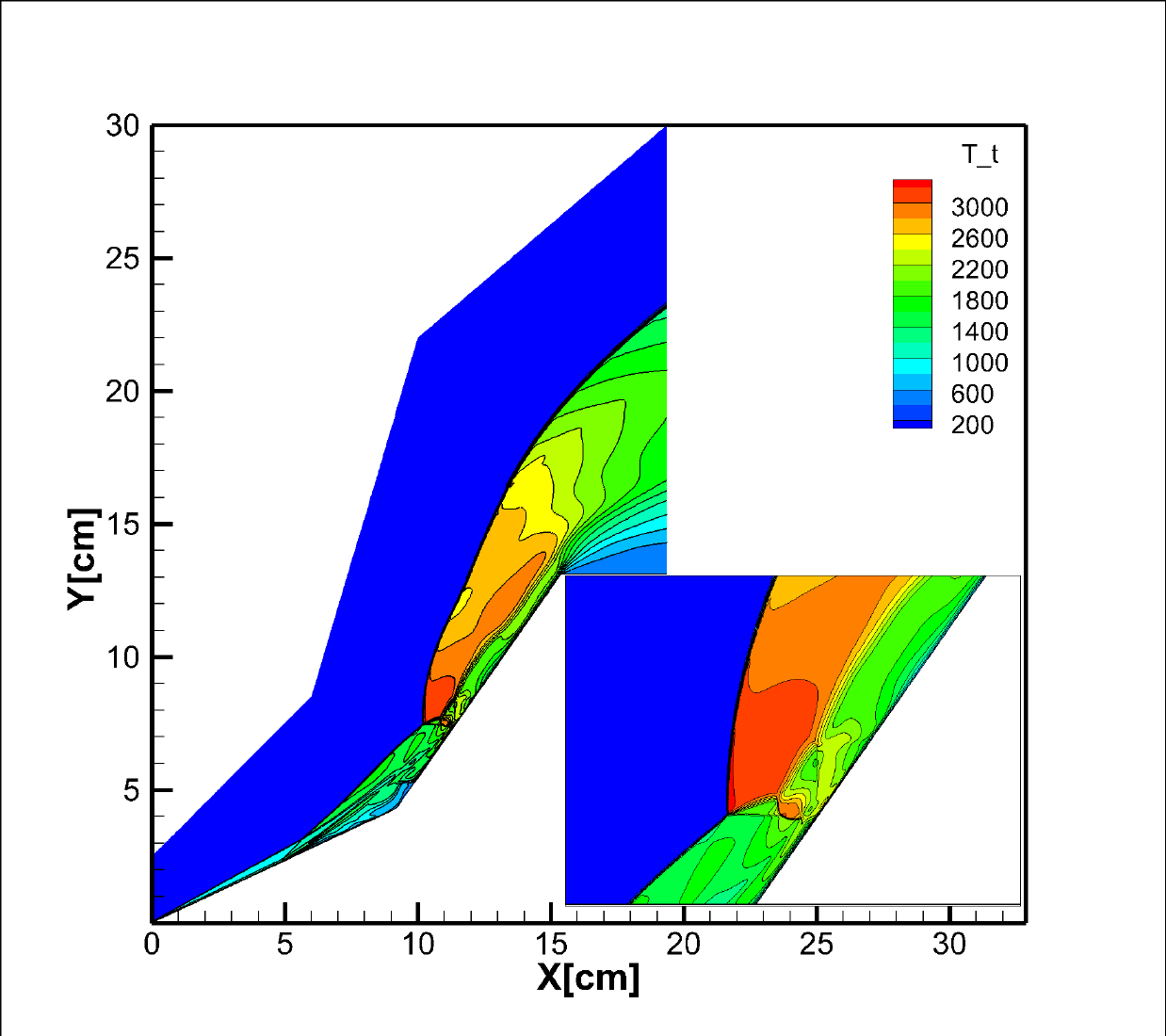}}
	\subfigure[rotational temperature]{\includegraphics[viewport=20 10 540 500,clip=true,width=0.485\textwidth]{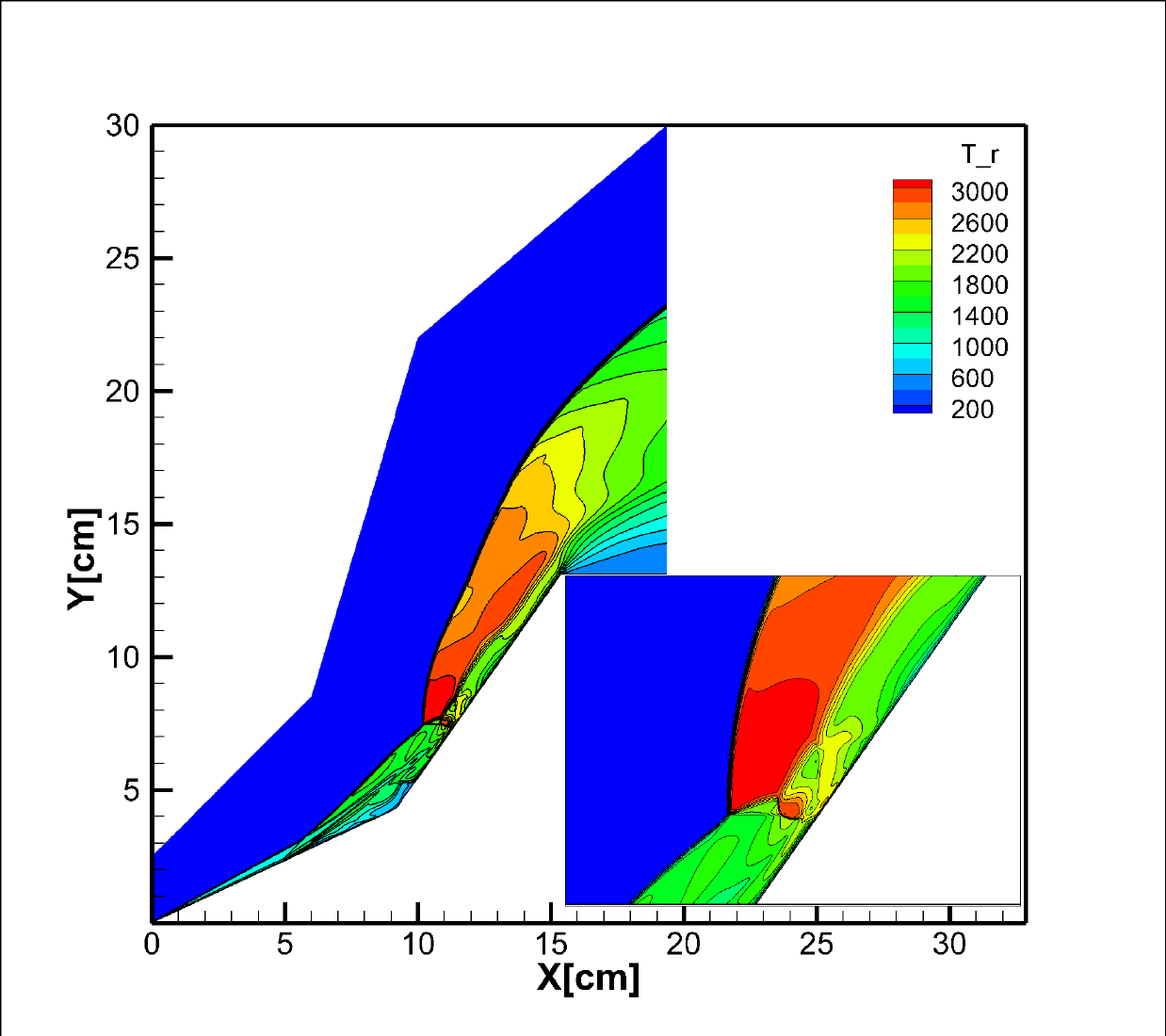}}
	\subfigure[vibrational temperature]{\includegraphics[viewport=20 10 540 500,clip=true,width=0.485\textwidth]{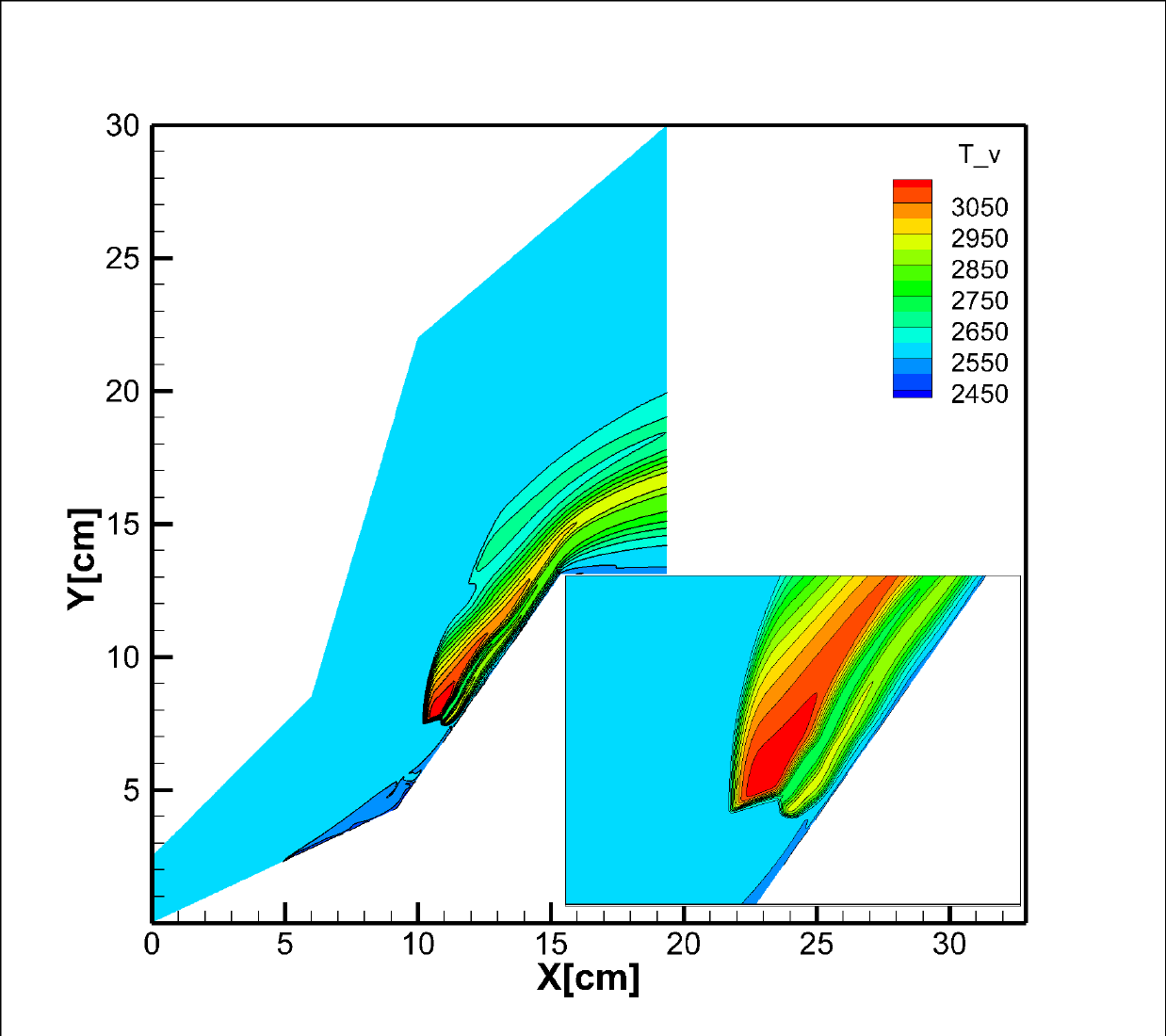}}
	\caption{Temperature contours (zoomed near the triple point to show the structure of oblique, bow and transmitted shocks) in the Run28 case. }  
	\label{fig:Run28flowfieldandtemperature}
\end{figure}

An in-house code based on the finite volume solver with multi-block parallel processing and implicit LU-SGS method \citep{LUSGS} is used in the numerical simulation. The inviscid flux scheme is the well-used AUSMPW+ scheme  \citep{KIM200138}, while the viscous term is discretized by a central difference method. 
After a grid-convergence study, the final spatial grid of the size $800 \times 500 \times 30$ is used, which contains 800 grids along the surface, 500 grids in the perpendicular direction of the surface, and 30 grids in the circumferential direction because of the axial-symmetry property of the configuration. 
Following the suggestion of \cite{Nompelis2004}, the distance of first grid perpendicular to the wall is set as $0.00001$~m, which leads to a cell Reynolds number of order 1. A localized mesh refinement around the separation zone and reattachment point is conducted to capture the complex flow structure.

 \begin{figure}
	\centering
	\subfigure[Run28]{\includegraphics[viewport=20 10 540 500,clip=true,width=0.4\textwidth]{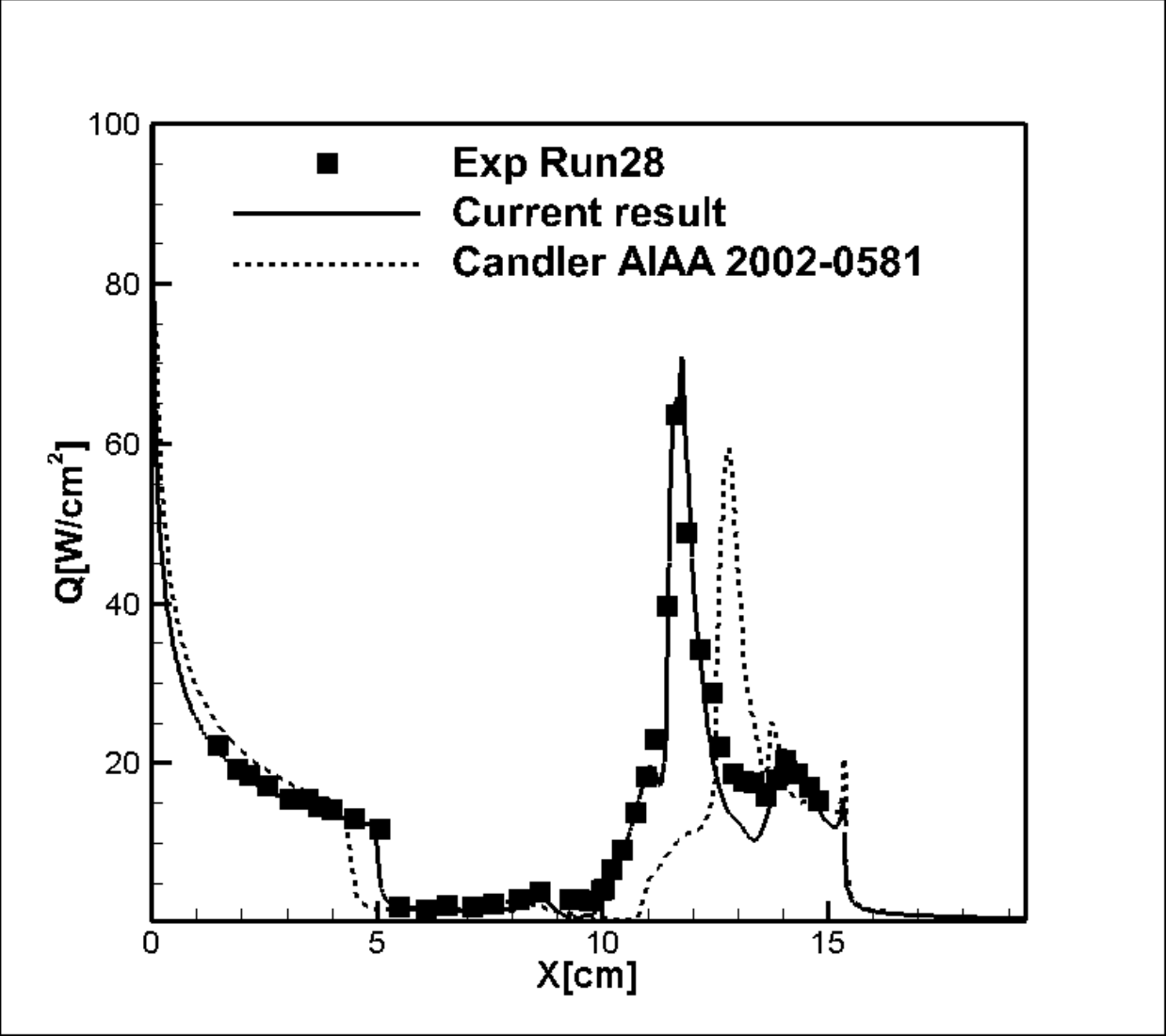}}
	\subfigure[Run35]{\includegraphics[viewport=20 10 540 500,clip=true,width=0.4\textwidth]{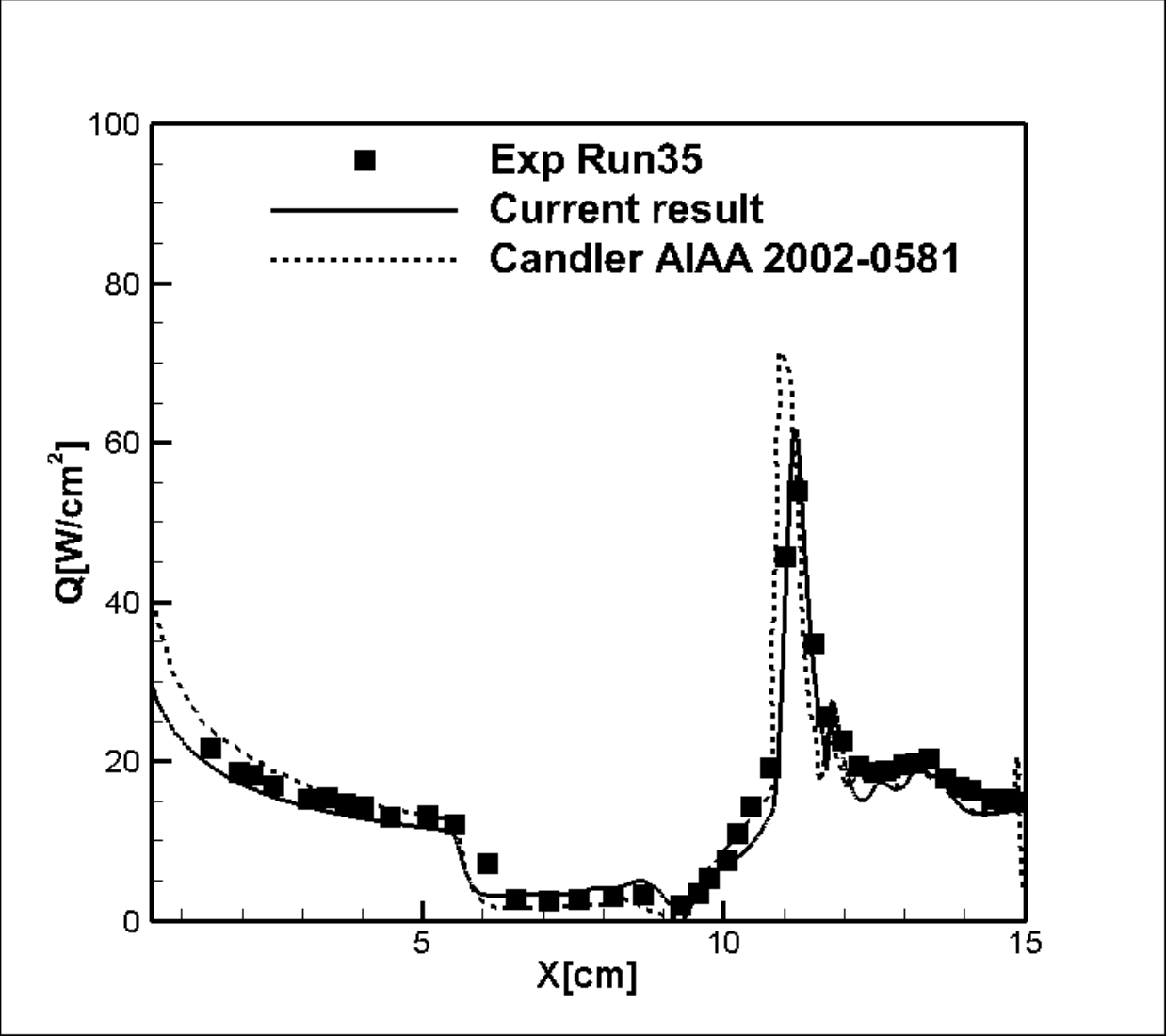}}\\
	\subfigure[Run42]{\includegraphics[viewport=20 10 540 500,clip=true,width=0.4\textwidth]{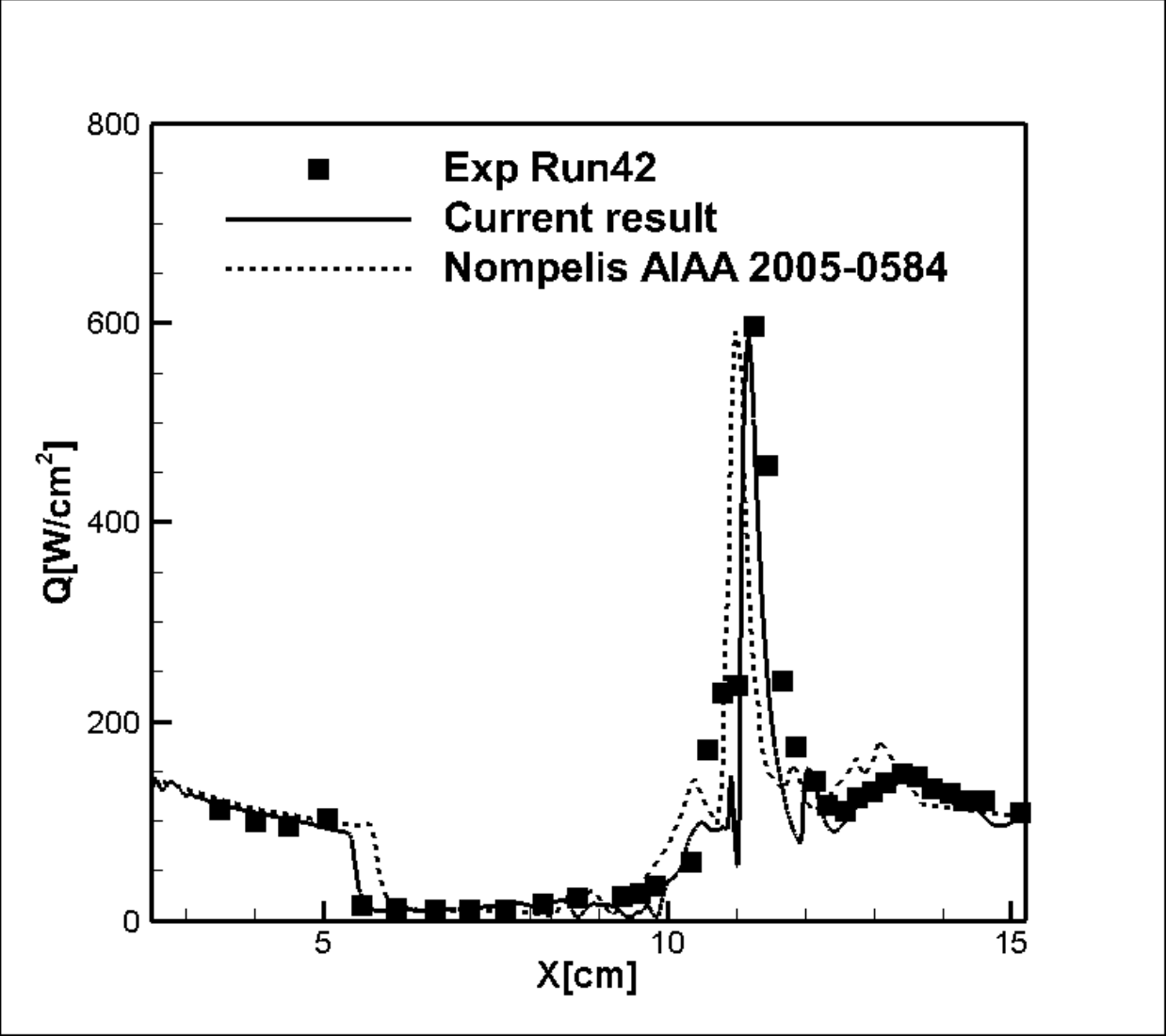}}
	\subfigure[Run46]{\includegraphics[viewport=20 10 540 500,clip=true,width=0.4\textwidth]{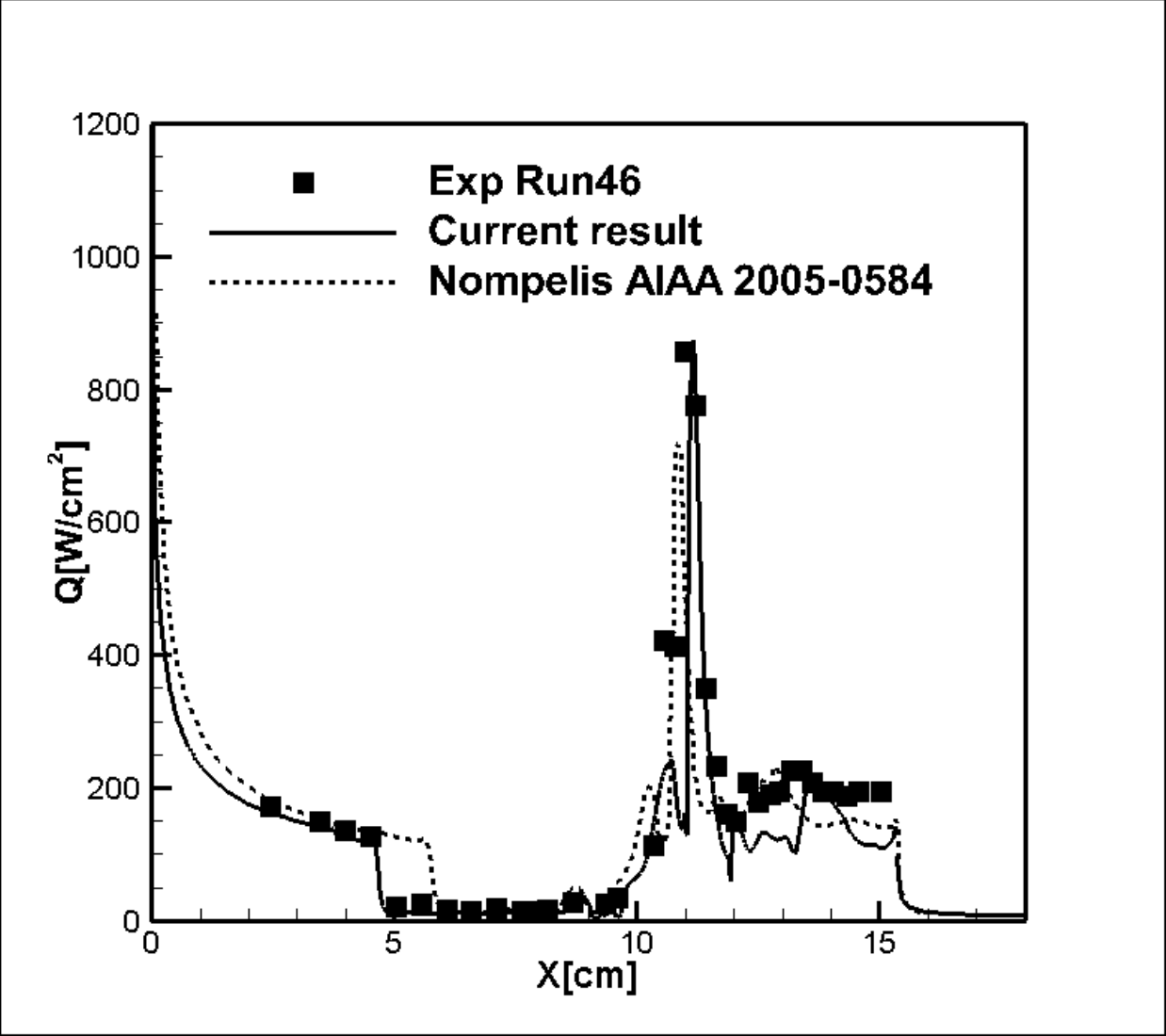}}
	\caption{The distribution of heat flux along the cone surface for four different cases with the inflow conditions summarized in table~\ref{tab:fivecasecondition}.}
	\label{fig:Run2835804246surfaceheatflux}
\end{figure}

We first consider the low enthalpy Run28  case, where the total enthalpy per unit mass is $ 3.4~\mathrm{MJ} / \mathrm{kg}$. 
The computed contours of Mach number for Run28 is displaced in figure~\ref{fig:dc_config}(\textit{b}). As shown in figure \ref{fig:Run28flowfieldandtemperature}(\textit{a}) and (\textit{b}),
the translational and rotational temperature contours are nearly the same. 
This is because that the inflow condition falls in the near-continuum regime, and the rotational collision number $Z_r$ is small so that the translational and rotational energies can quickly reach equilibrium. 
As for the vibrational temperature, significant differences to the translational/rotational temperature could be seen in figure \ref{fig:Run28flowfieldandtemperature}(\textit{c}). 
The thermodynamic non-equilibrium comes from {the large value of the vibrational collision number $Z_v$} \citep{Knight2018}, which is about $10^3\sim10^4$ as per \eqref{vibrational_collision_number}. 
As a result, the vibrational temperature remains unchanged until the interaction of separated shock and detached shock. 
After the interaction, the vibrational temperature climbs to a peak value around 3100~K. 
Due to the long residence time in the separation zone, the vibrational temperature falls to a level around 2400~K. The similar is true in the Run35 case.

The simulated heat flux along the surface of double cone for the Run 28 case is shown in figure~\ref{fig:Run2835804246surfaceheatflux}(\textit{a}).
It could be seen that the heat flux peak as well as the length and location of the separation zone from our result agree better with experiment data than that from \citet{AIAA20020581} using the original Eucken factor~\eqref{Euken_2}. 
To be specific, our method can predict the position of steep drop of heat flux more accurately which indicates a more accurate starting location of separation zone. 
Moreover, our method can improve the predicting accuracy for the position of heat flux peak. Because the distance between the steep drop and peak represents the size of separation zone, the numerical results indicate that our NSF model and boundary conditions makes a big improvement in capturing the separation zone.

For the low total enthalpy Run35 case in figure \ref{fig:Run2835804246surfaceheatflux}(\textit{b}), our numerical results also predict the heat flux well, in terms of the peak value and position. We also note that the peak heat transfer obtained by \citet{AIAA20020581} is slightly larger than the experimental data.

The third and fourth test cases are the Run42 and Run46, respectively, which 
have much higher stagnation enthalpies than the previous two cases. Therefore, these two cases have a significant high-temperature gas effect. This could be a challenge in the numerical simulation of hypersonic SWBLI flows as reported by \citet{Nompelis2005}. The simulated heat flux using the corrected Eucken factors~\eqref{ft_fint_fv} and the modified temperature-jump boundary conditions~\eqref{eq:temperature_jump_current} is shown in figure~\ref{fig:Run2835804246surfaceheatflux}(\textit{c}) and (\textit{d}). An overall good agreement are seen. The abrupt decrease of heat flux at $x\approx5$~cm showing the starting of flow separation moves further upstream than that from \citet{Nompelis2005}, while the peak of heat flux showing the reattachment point moves downstream. The location of the abrupt decrease and peak for both cases match well with experiment data. 

\section{Influence of different heat conductivities}\label{sec:Euckendiscussion}

Figure~\ref{fig:Run2835804246surfaceheatflux} show that our method improves the accuracy in the prediction of surface heat flux, especially for the location of steep drop and peak value.
Considering that our main modifications come from a curve-fitting of Eucken factors and the temperature-jump boundary conditions, in this section, attention would be made to the role of Eucken factors. To be specific, numerical results using the original and corrected Eucken factors, as given by~\eqref{Euken_2} and~\eqref{ft_fint_fv}, respectively, will be compared; the latter is widely used in the simulation of thermochemical nonequilibrium flows. To make a fair comparison, the same temperature-jump conditions~\eqref{eq:temperature_jump_current} are used. 

\subsection{The Run28 case}

The Run28 case is analysed as a representative of low enthalpy flow.
From figure \ref{fig:Run2835804246surfaceheatflux}(\textit{a}), it is seen that the difference of the surface heat flux between our method and \cite{AIAA20020581} are the position of steep drop and peak. 
Considering the flow pattern of shock wave and boundary layer interaction in figure~\ref{fig:dc_config}(\textit{b}), other flowfields computing from the original and corrected Eucken factors are compared.

\begin{figure}
  \centering
   \subfigure[]{\includegraphics[viewport=20 20 540 500,clip=true, width=0.4\textwidth]{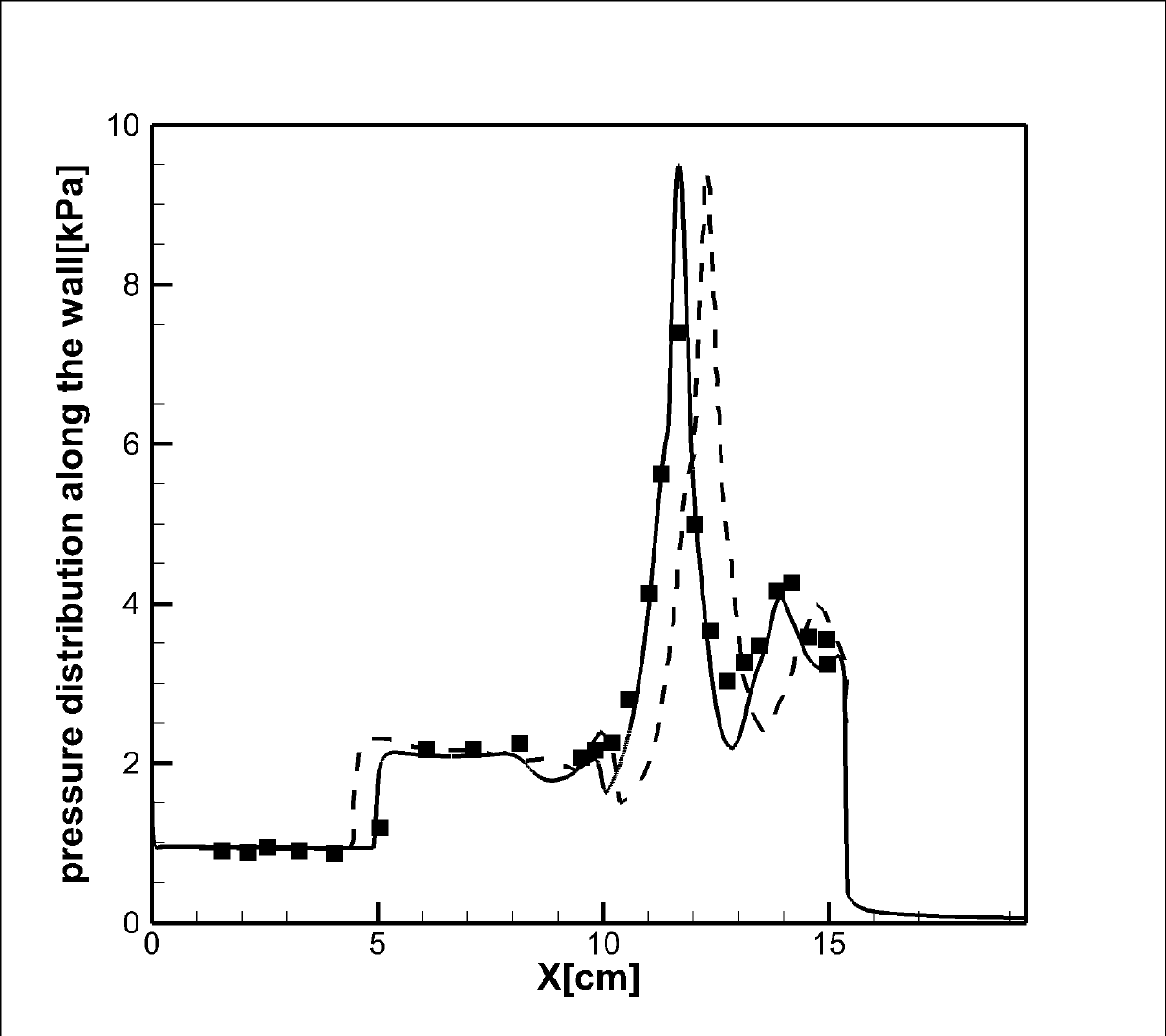}}
  \subfigure[]{\includegraphics[viewport=20 20 540 500,clip=true, width=0.4\textwidth]{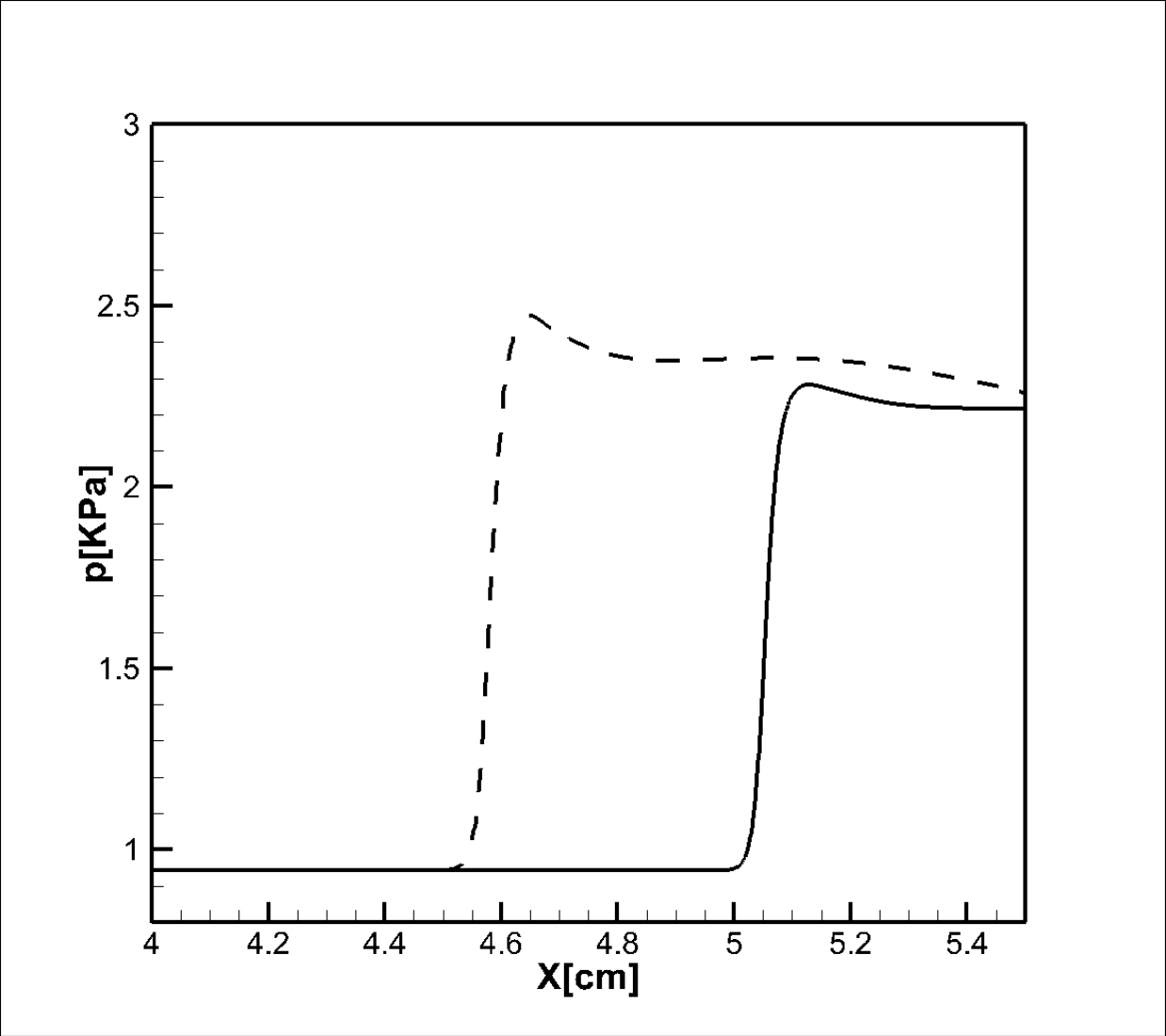}}
  \subfigure[]{\includegraphics[viewport=20 20 540 500,clip=true, width=0.4\textwidth]{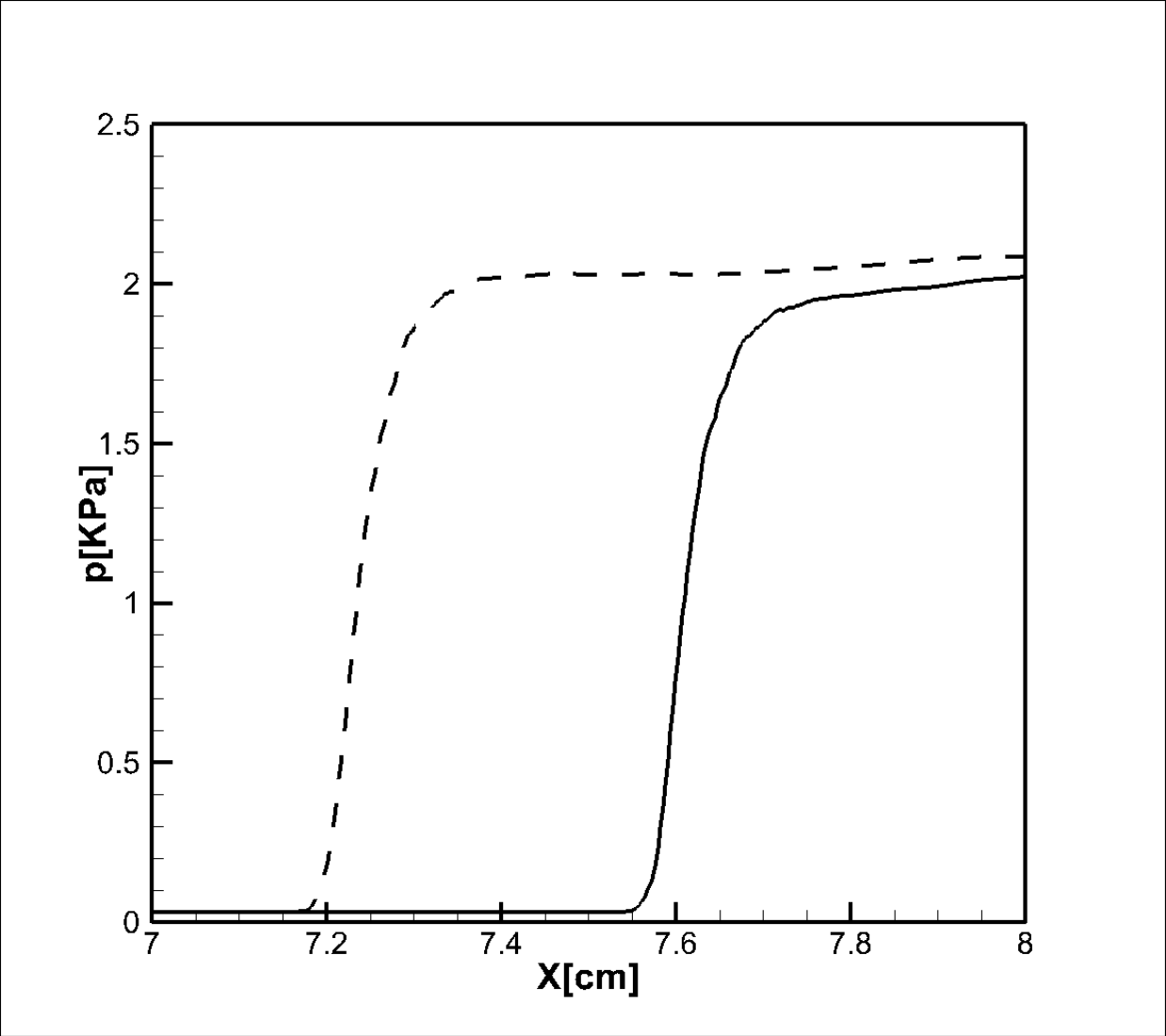}}
  \subfigure[]{\includegraphics[viewport=20 20 540 500,clip=true, width=0.4\textwidth]{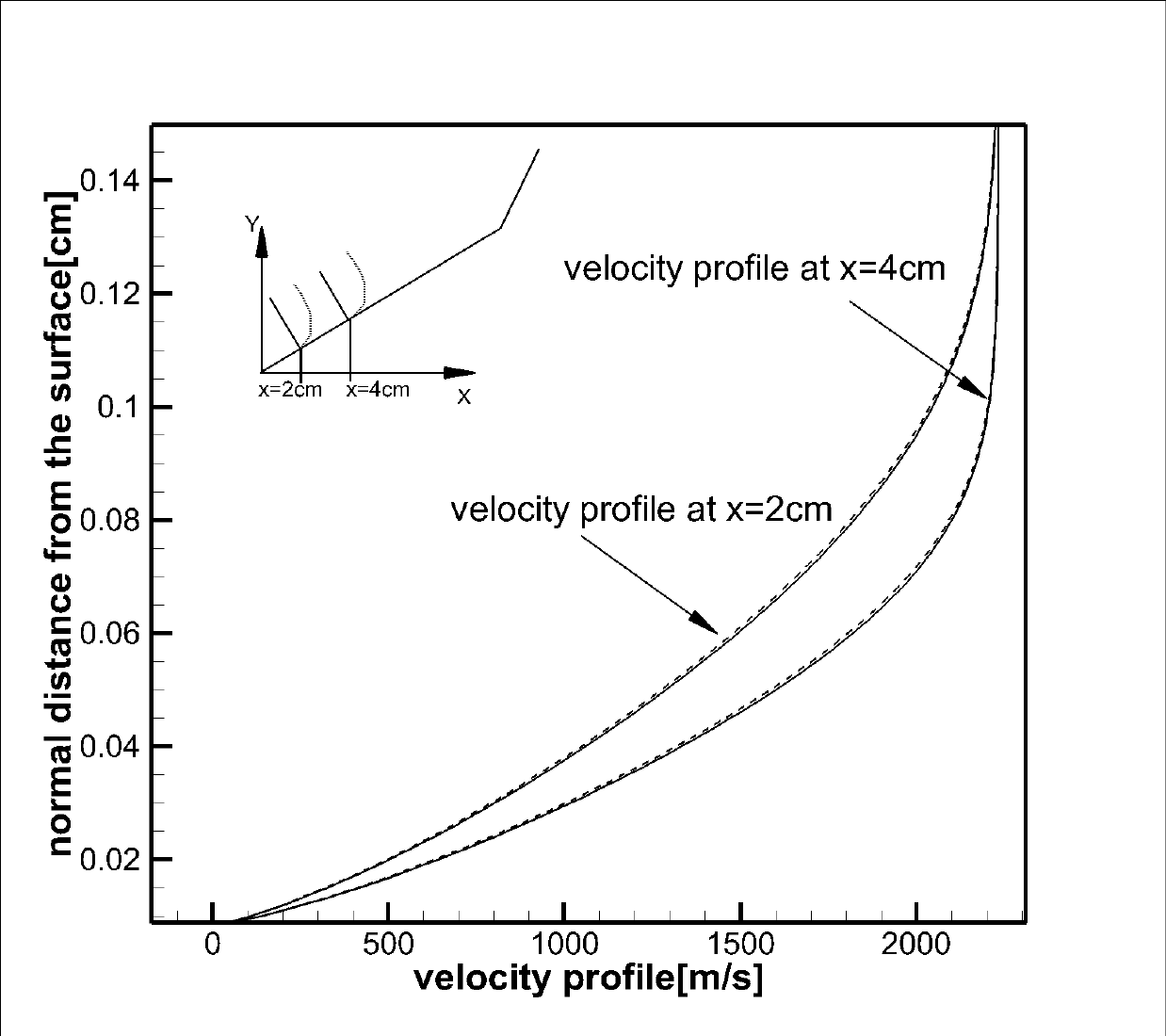}}
  \caption{The Run28 case. Comparison of the pressure distribution along the (a)  surface (b) streamline started from  $x=-\infty, y=1$~cm, and (c) streamline started from $x=-\infty, y=5$~cm. (d) Comparison of the  horizontal velocity distribution at $x=2$~cm and $x=4$~cm. Solid and dashed lines: results from the corrected
  	\eqref{ft_fint_fv} and original \eqref{Euken_2} Eucken factors, respectively. Squares:  experimental data. }
\label{fig:Run28adversepressurebijiao}
\end{figure}

The pressure profile along the surface is compared in \ref{fig:Run28adversepressurebijiao}(\textit{a}), where the numerical result obtained from the NSF equations with the corrected Eucken factors agrees well with the experimental data, while that of the original Eucken factors predicts an early pressure rise around $x=5$~cm and late pressure peek position around $x=12$~cm. The pressure profiles along two streamlines (one starts from $x=-\infty,y=1$~cm that goes through the oblique shock, and the other from $x=-\infty,y=5$~cm that goes through the separated shock) are also  shown in figure \ref{fig:Run28adversepressurebijiao}(\textit{b}) and (\textit{c}). It could be seen that the pressure obtained from the NSF equations with the corrected Eucken factor is lower than that form the original Eucken factor. 
This is because the shock wave compression converts the kinetic energy into the  internal energy, and the larger translational heat conductivity in \eqref{Euken_2} would allocate more transformed energy into the translational mode, and thus leads to larger gas pressure.
The adverse pressure gradient (at $x\approx5$~cm) at the wall computed from the original Eucken factor is larger than that of the corrected Eucken factor. 
However, the velocity distribution along the vertical direction of the surface (starting from the surface at $x=2$~cm and $x=4$~cm) remains the same, see  figure \ref{fig:Run28adversepressurebijiao}(\textit{d}).
With the same velocity distribution and smaller adverse pressure gradient,
the beginning location of the separation bubble (around $x=4.9$~cm) computed using the corrected Eucken factor is delayed than that from the original one (around $x=4.5$~cm).
The boundary layer separation weakens the heat transfer between the gas and surface. As a result, an abrupt decrease of heat flux is observed around this separation.
The delay of separation finally results in the delay of steep drop of surface heat flux as shown in figure \ref{fig:Run2835804246surfaceheatflux}(\textit{a}). 

\begin{figure}
	\centering
	\subfigure[]{\includegraphics[viewport=20 20 540 500,clip=true, width=0.4\textwidth]{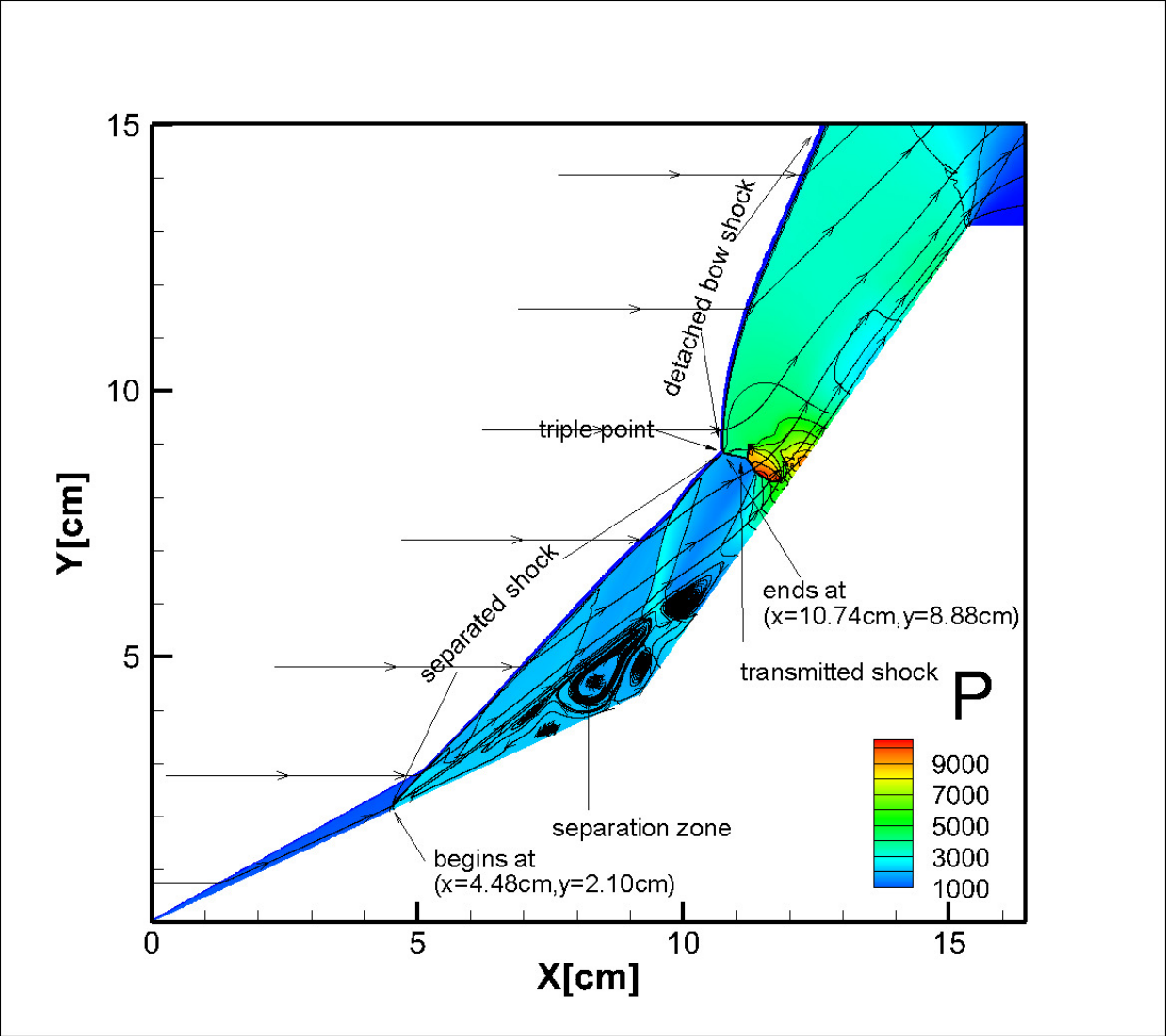}}
	\subfigure[]{\includegraphics[viewport=20 20 540 500,clip=true, width=0.4\textwidth]{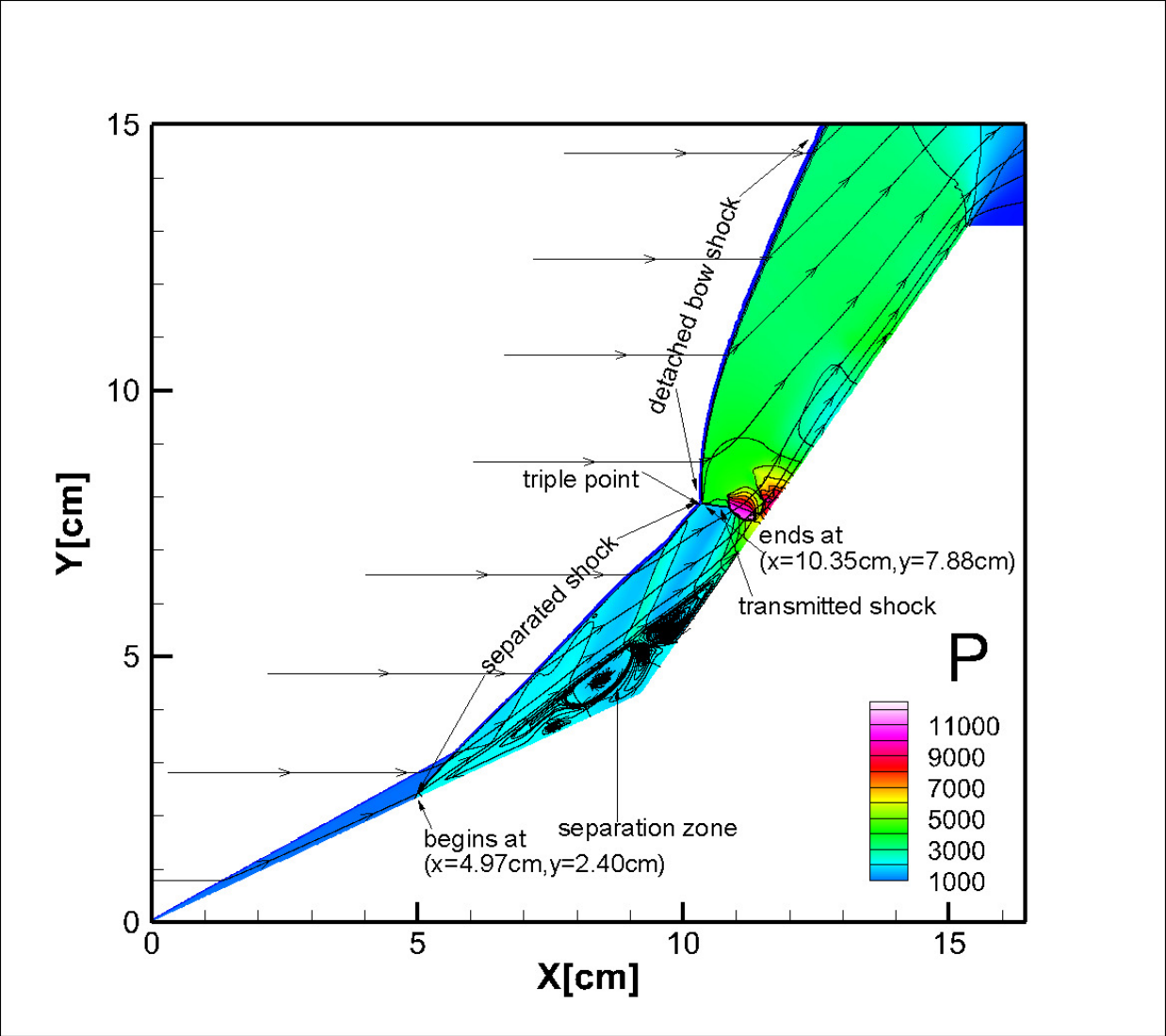}}\\
	\subfigure[]{\includegraphics[viewport=20 20 540 500,clip=true, width=0.4\textwidth]{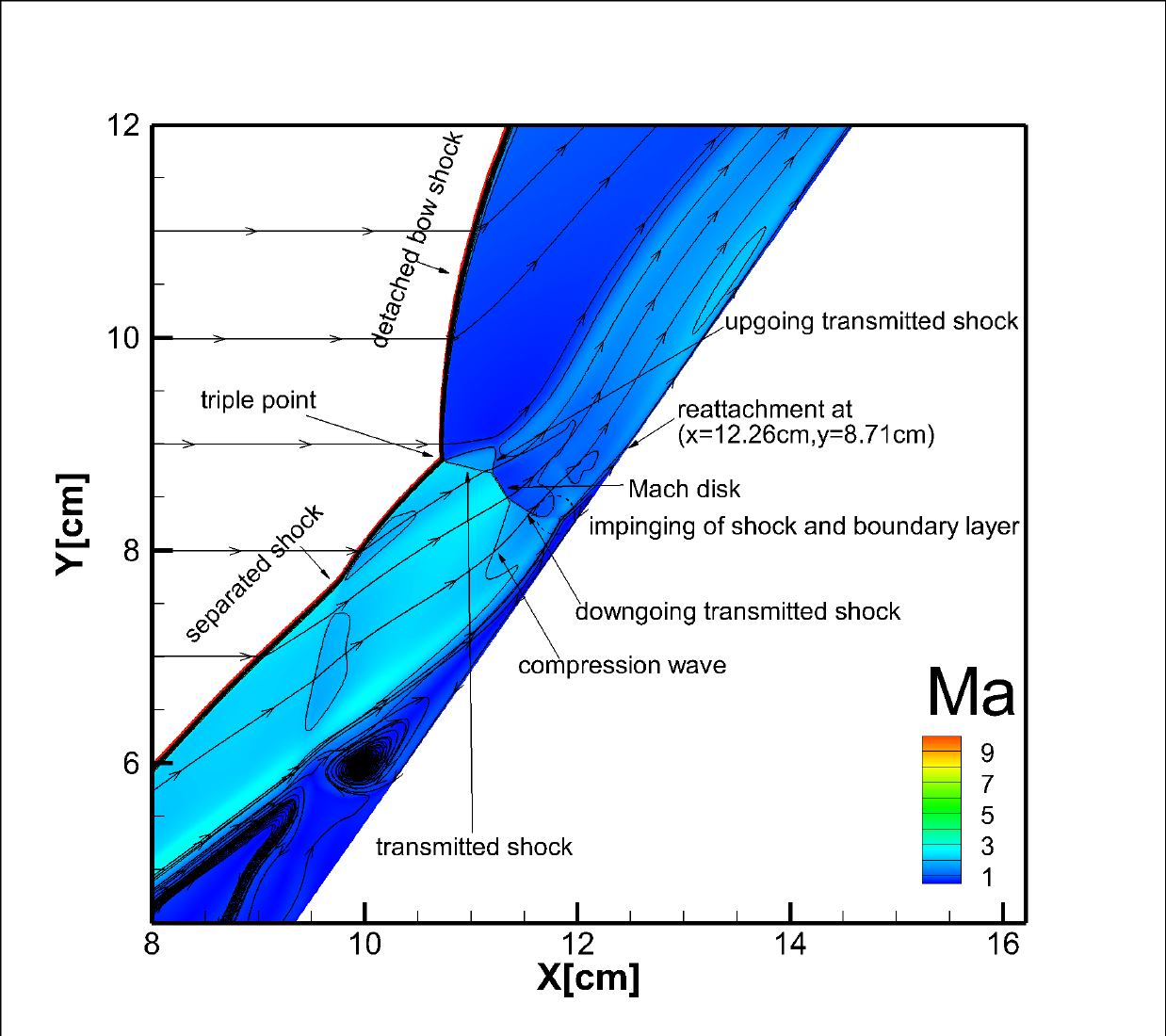}}
	\subfigure[]{\includegraphics[viewport=20 20 540 500,clip=true, width=0.4\textwidth]{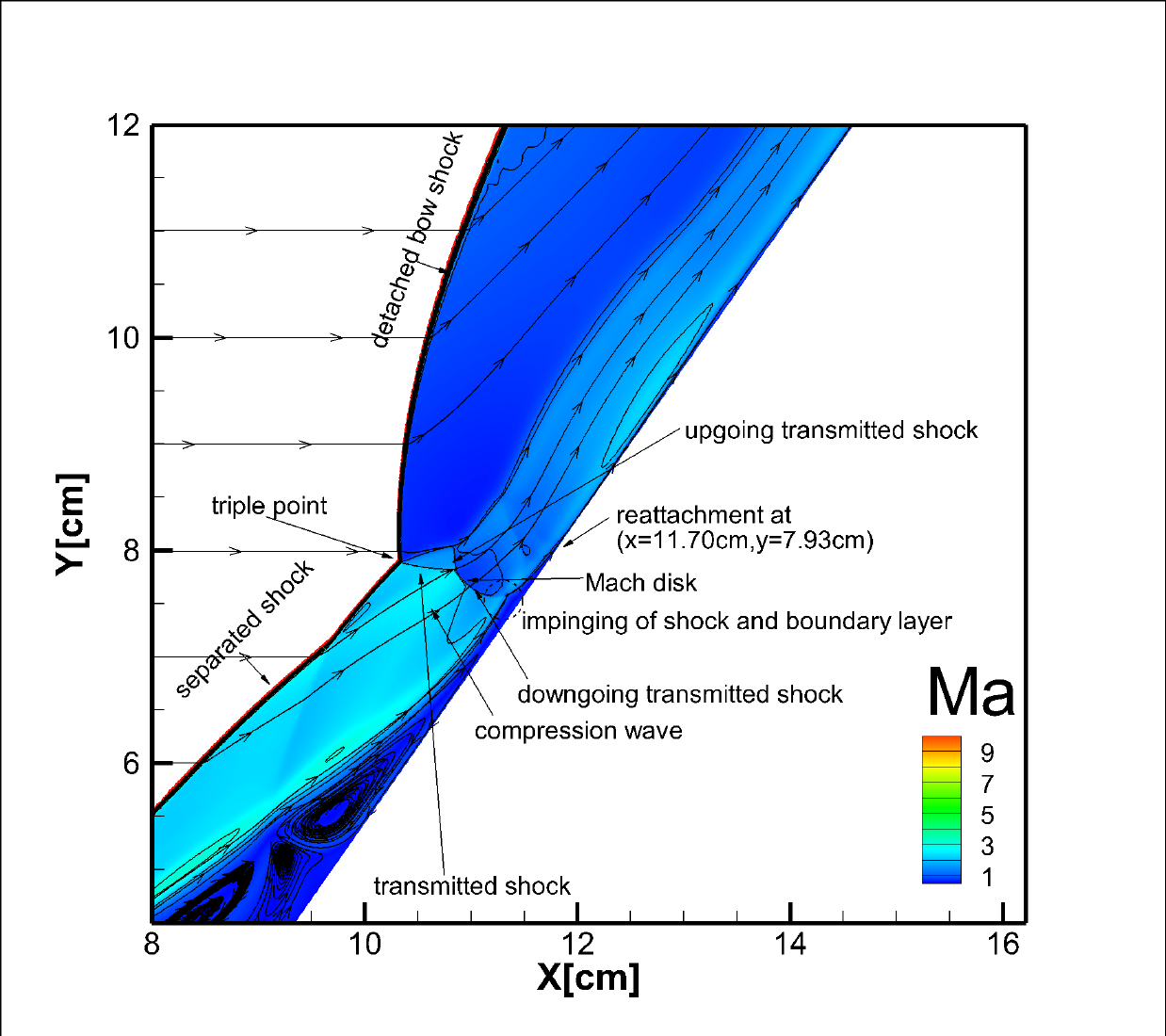}}\\
	\subfigure[]{\includegraphics[viewport=20 20 540 500,clip=true, width=0.4\textwidth]{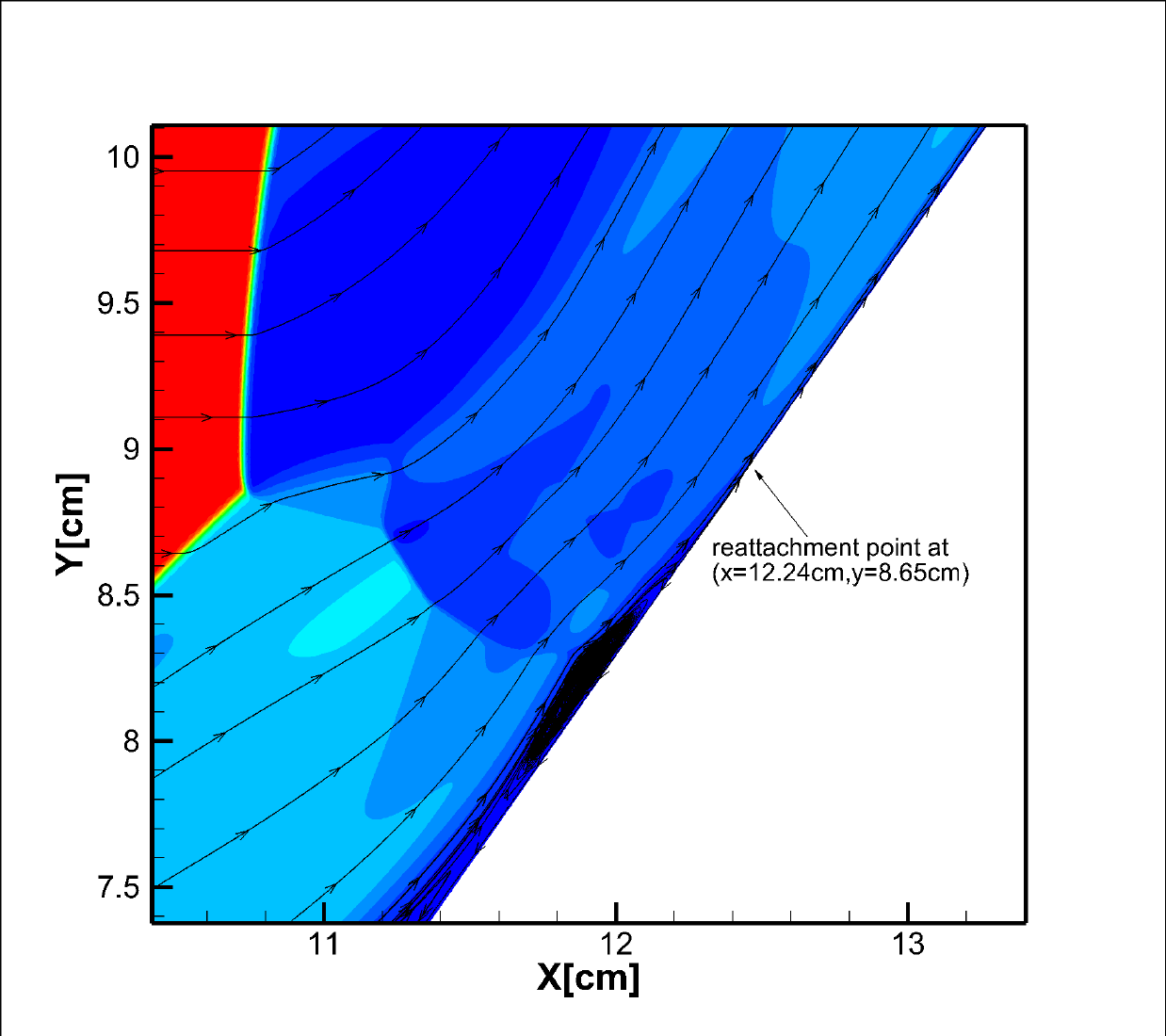}}
	\subfigure[]{\includegraphics[viewport=20 20 540 500,clip=true, width=0.4\textwidth]{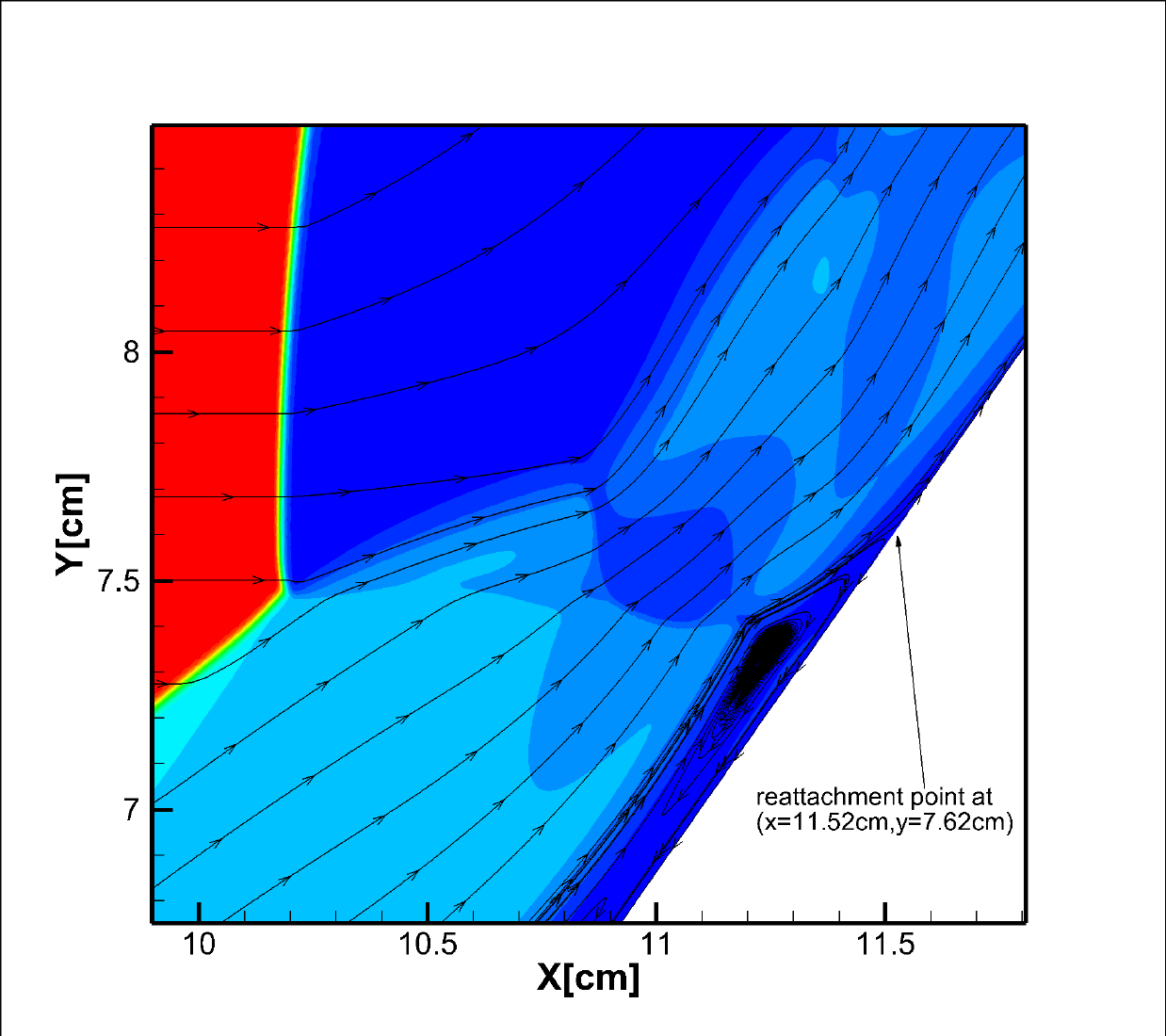}}
	\caption{Comparison of separated shock wave (first row), triple point and transmitted shock wave (second row), and reattachment point (third row) between the original (left column) and corrected (right column) Eucken factors, for the Run28 case.}
	\label{fig:Run28separatedshockbijiao}
\end{figure}

The beginning and ending positions of the separated shock waves, obtained from the original and corrected Eucken factors, are shown respectively in figure \ref{fig:Run28separatedshockbijiao}(\textit{a}) and (\textit{b}) for comparison. When the original Eucken factor is used in the NSF equations, the beginning position of the separated shock wave is around $(x,y)=(4.48,2.10)$~cm, while the ending position is around $(10.74,8.88)$~cm. Note that the ending position is also the position of the triple point which is a critical flow structure~\citep{wuziniu2010,wuziniu2013}. 
The inflow conditions and experiment geometry need to be carefully chosen to make sure that the triple point is located at a fixed position. Otherwise, it can move at a nearly constant speed in the transition from the regular reflection to the Mach reflection~\citep{wuziniu2011}.
In the current research, the triple point is located at a fixed position.  When the modified Eucken factor is used, the beginning and ending positions of the  separated shock wave are around $(x,y)=(4.97,2.40)$~cm and $(10.35, 7.88)$~cm, respectively. 
Therefore, a shorter distance between the triple point and surface is computed from the corrected Eucken factor. 

Different position of the triple point leads to different transmitted shock waves, which are compared in figure~\ref{fig:Run28separatedshockbijiao}(\textit{c}) and (\textit{d}) for the original and corrected Eucken factors, respectively. 
The transmitted shock wave after the triple point forms three structures: an upgoing transmitted shock, a Mach disk, and a downgoing transmitted shock. 
The downgoing transmitted shock orients toward the surface. 
The flow above the separation zone accelerates towards the surface, resulting in a reduction of the thickness of local boundary layer. 
Eventually, different impinging regions between the downgoing transmitted shock and boundary layer results in different reattachment points, which are shown in figure~\ref{fig:Run28separatedshockbijiao}(\textit{e}) and (\textit{f}). While the original Eucken factor leads to a position of $x=12.24$ cm in the reattachment point, the corrected Eucken factor in the NSF equation predicts a shorter distance between the triple point and surface, making the reattachment earlier at about $x=11.52$ cm.

Reattachment enhances the heat transfer between the gas and surface, and the direct impinging between gas and surface could increase the heat flux to a maximum value.
Finally, this difference leads to different position of heat flux peak, as  figure~\ref{fig:Run2835804246surfaceheatflux}(\textit{a}) shows that, the corrected Eucken factor predicts a maximum heat flux at $x=11.75$ cm, while the original Eucken factor predicts at $x=12.28$ cm. It is seen that the maximum heat flux from our model is about 20\% higher than that from the original Eucken factors. 

\begin{figure}
  \centering
  {\includegraphics[viewport=20 20 540 500,clip=true, width=0.4\textwidth]{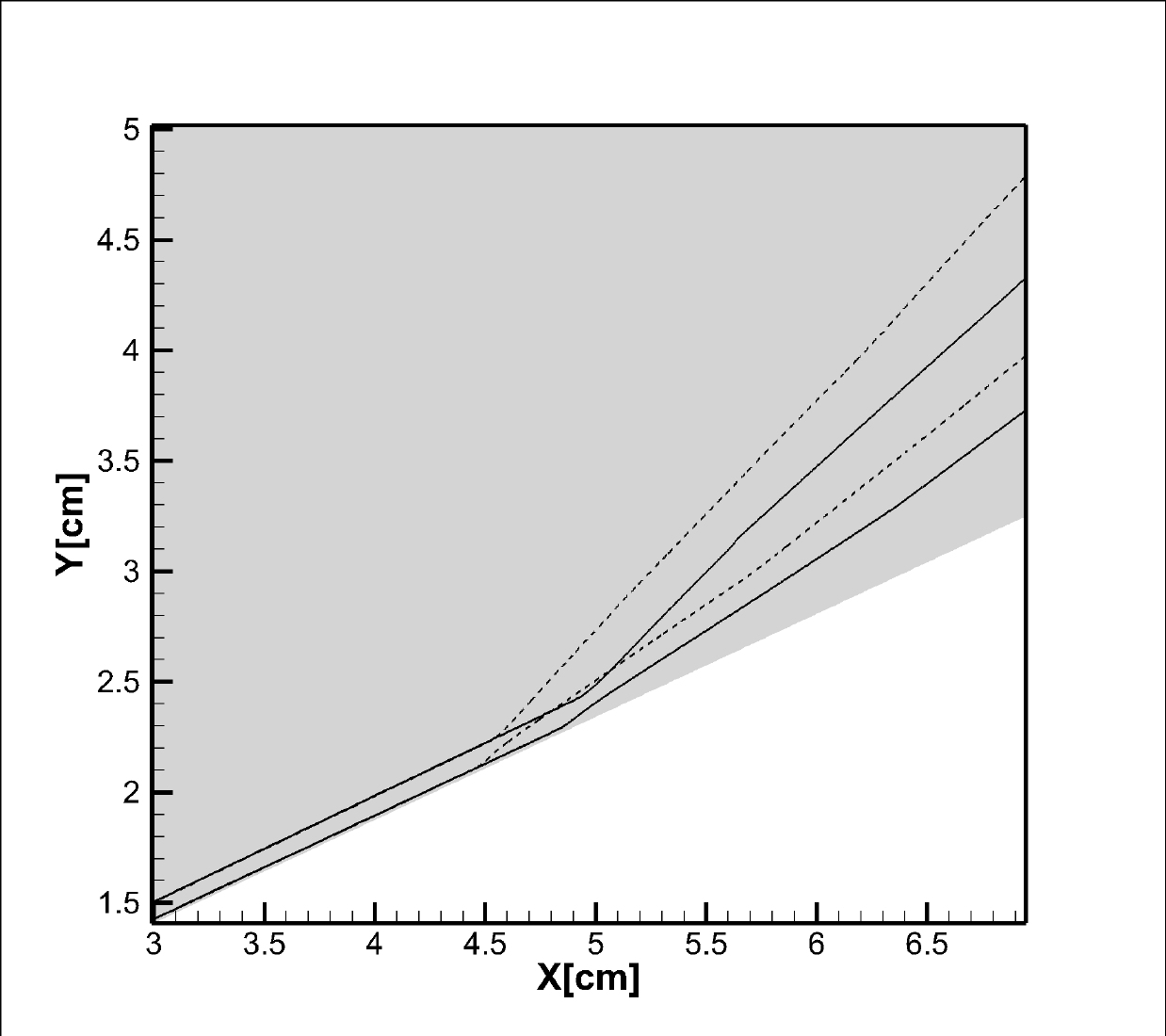}}
  \hskip 0.5cm
  {\includegraphics[viewport=20 20 540 500,clip=true, width=0.4\textwidth]{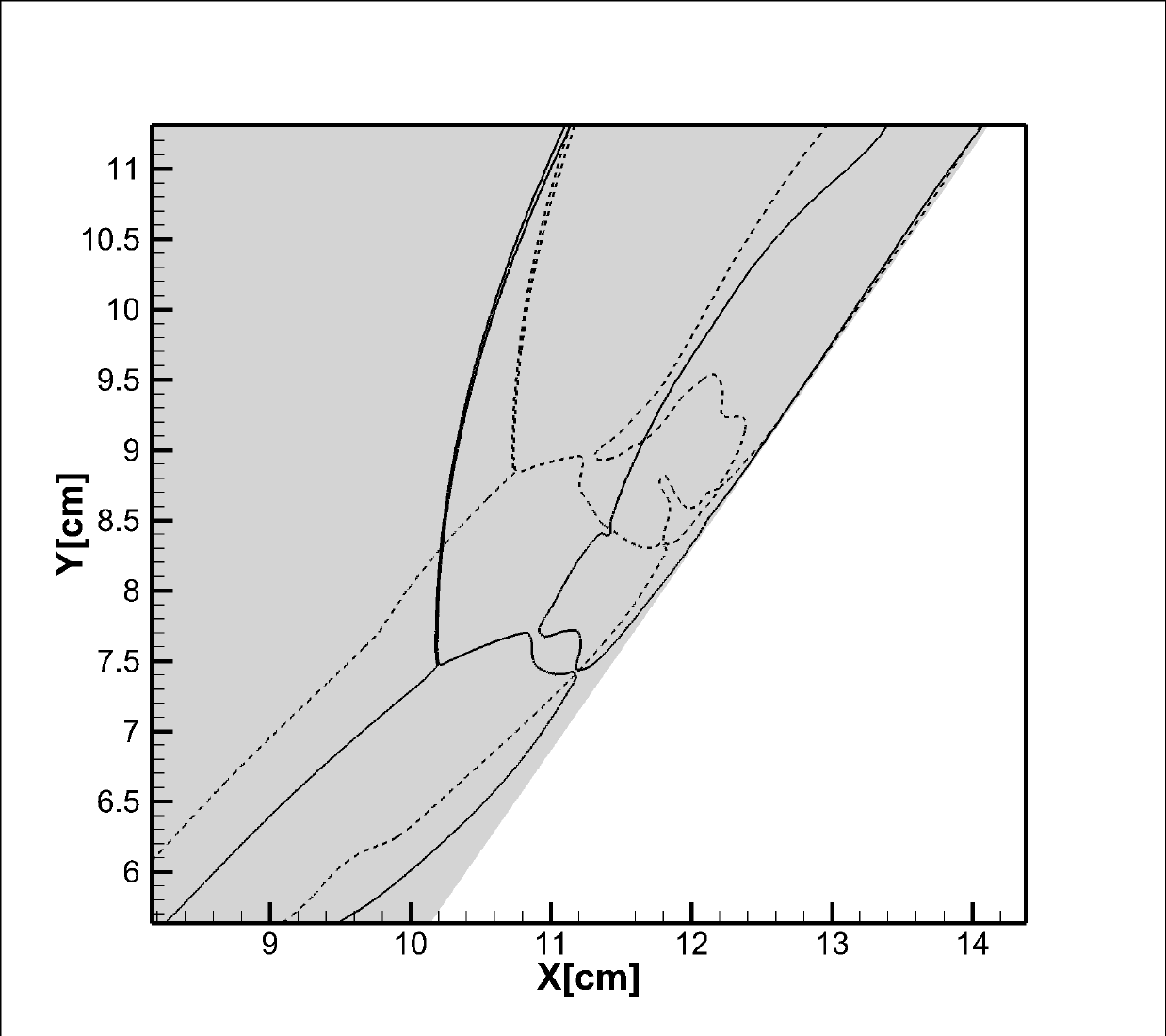}}
  \caption{
  	Comparison of the Mach number contour between different Eucken factors in the Run 28 case: (Left) in the shock wave interaction region, (Right) in the separation region. Solid and dashed lines: results from corrected and original Eucken factors, respectively. 
  }
\label{fig:Run28Malunkuobijiao}
\end{figure}

For better visualization, contours of the Mach number from the original and corrected Eucken factors are shown in figure \ref{fig:Run28Malunkuobijiao}. Different locations of the separation bubbles, triple points and transmitted shocks are clearly seen.

\subsection{The Run46 case}

The Run46 case is analysed as a representative of high enthalpy flow. First, it should be noted that the  
results in figure \ref{fig:Run2835804246surfaceheatflux}(\textit{d}) is convincing and encouraging of our modifications of the Eucken factors and temperature-jump boundary conditions. 
Although with significant high-temperature gas effect, numerical results of the heat flux agree well with the experiment data. An interesting observation is that, when the corrected Eucken factor is adopted in the NSF equation, the starting location of the separation moves upstream when compared to the original Eucken factor, and this trend is opposite to the low enthalpy Run28 case. Also, the peak heat flux obtained from the corrected Eucken factor moves downstream when compared to that from the original Eucken factor, which is again opposite to that in the low enthalpy flow.  

\begin{figure}
	\centering
	\subfigure[]{\includegraphics[viewport=20 10 540 500,clip=true,width=0.4\textwidth]{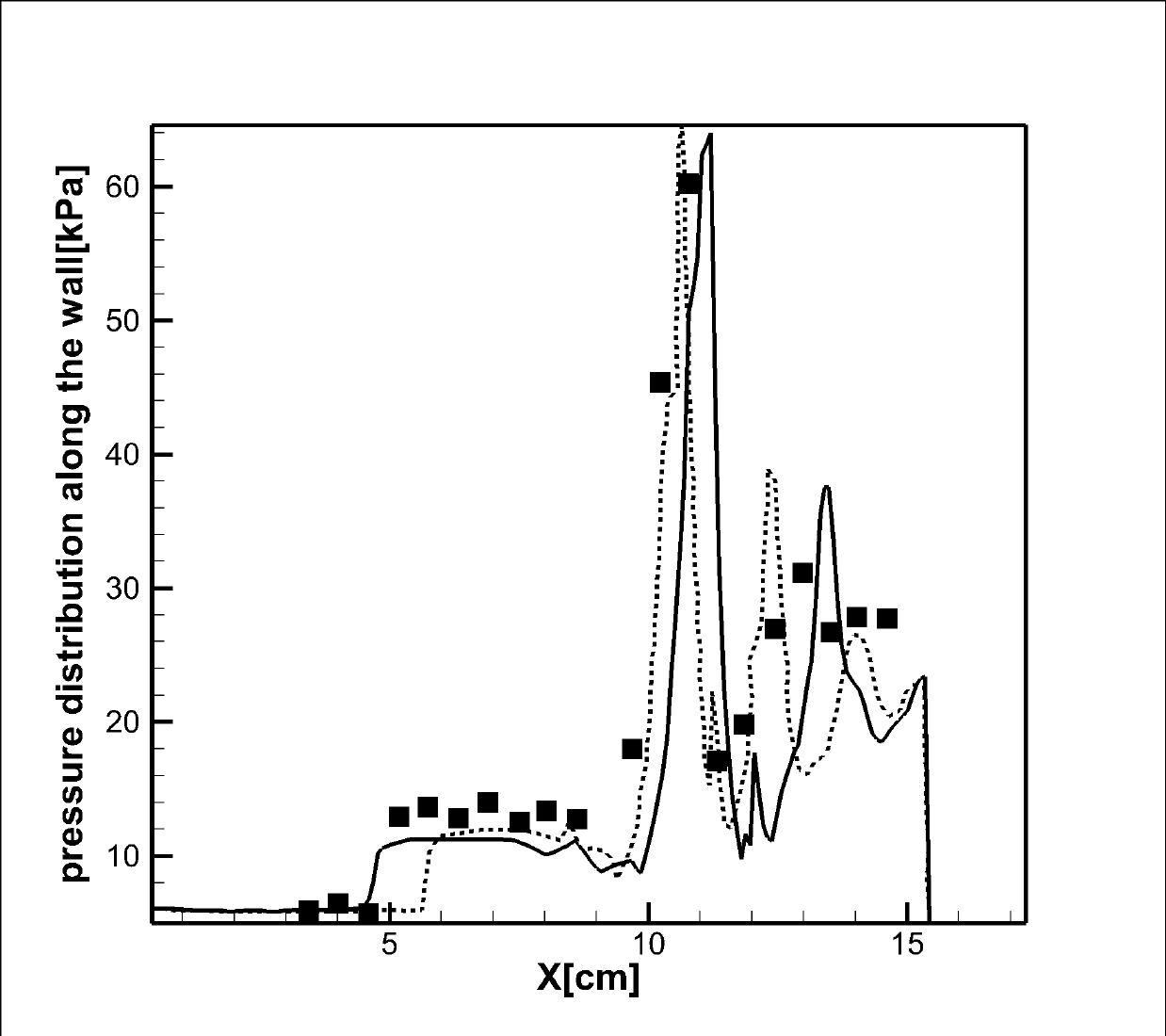}}
	\subfigure[]{\includegraphics[viewport=20 10 540 500,clip=true,width=0.4\textwidth]{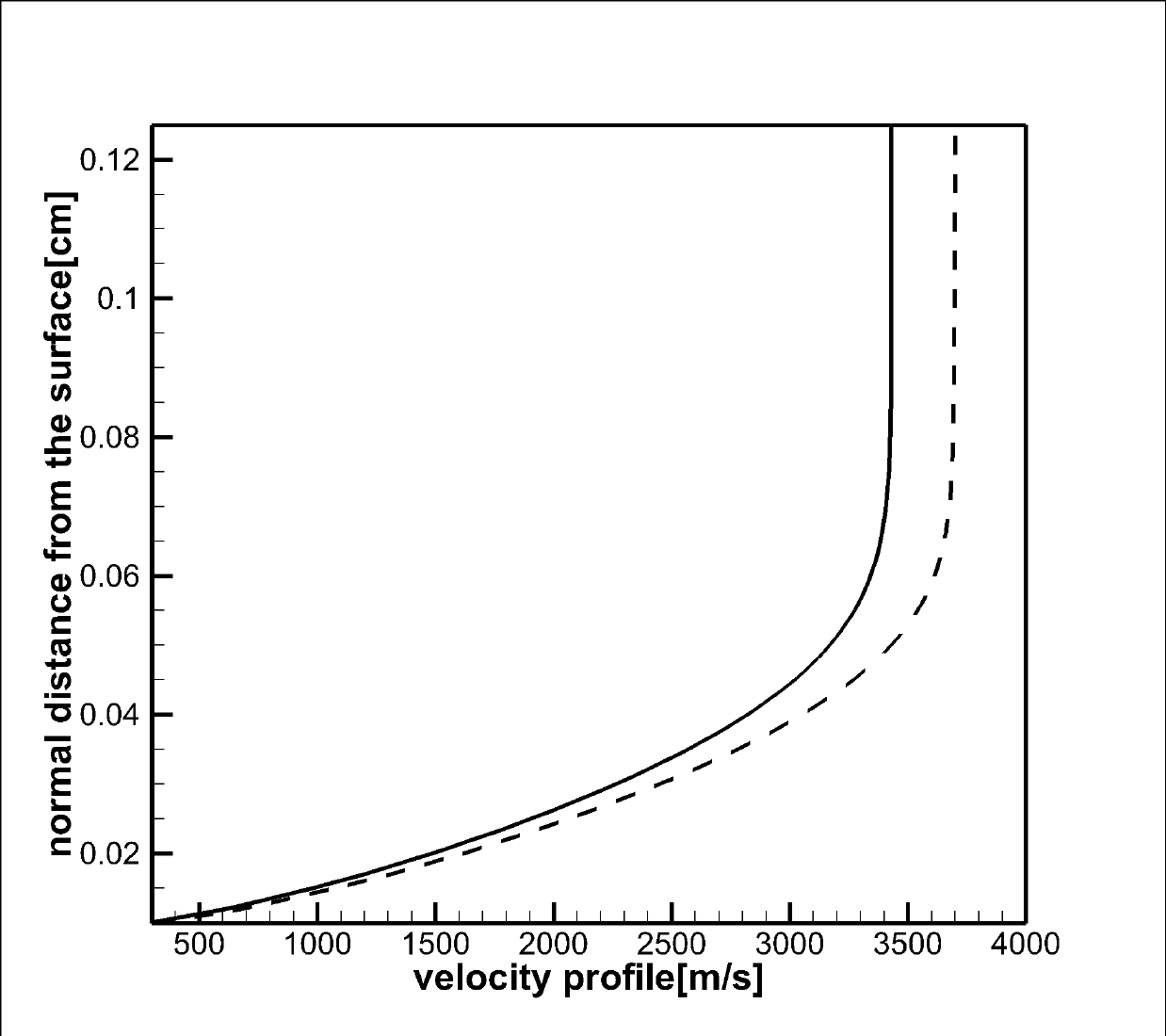}}
	\subfigure[]{\includegraphics[viewport=20 10 540 500,clip=true,width=0.4\textwidth]{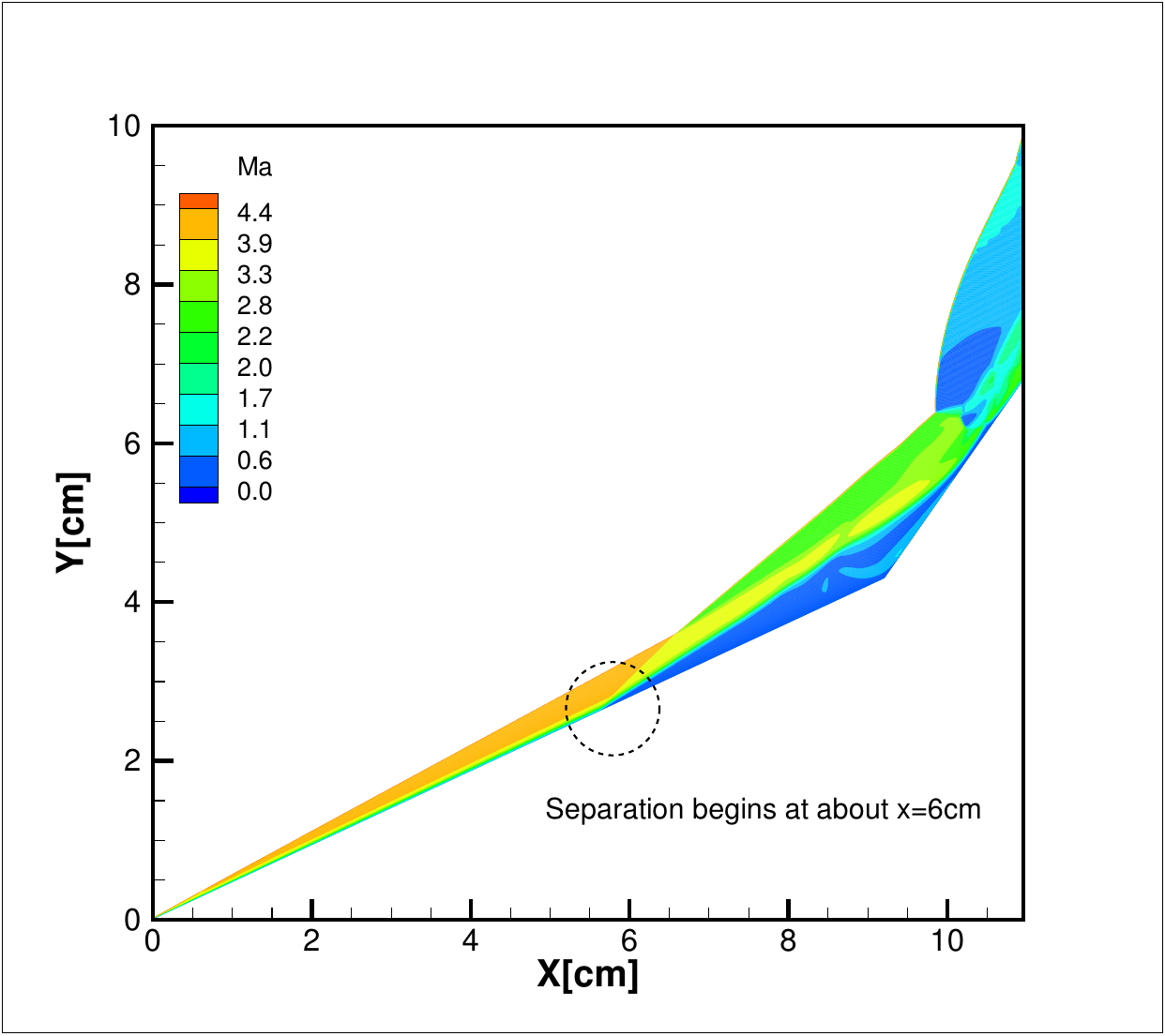}}
	\subfigure[]{\includegraphics[viewport=20 10 540 500,clip=true,width=0.4\textwidth]{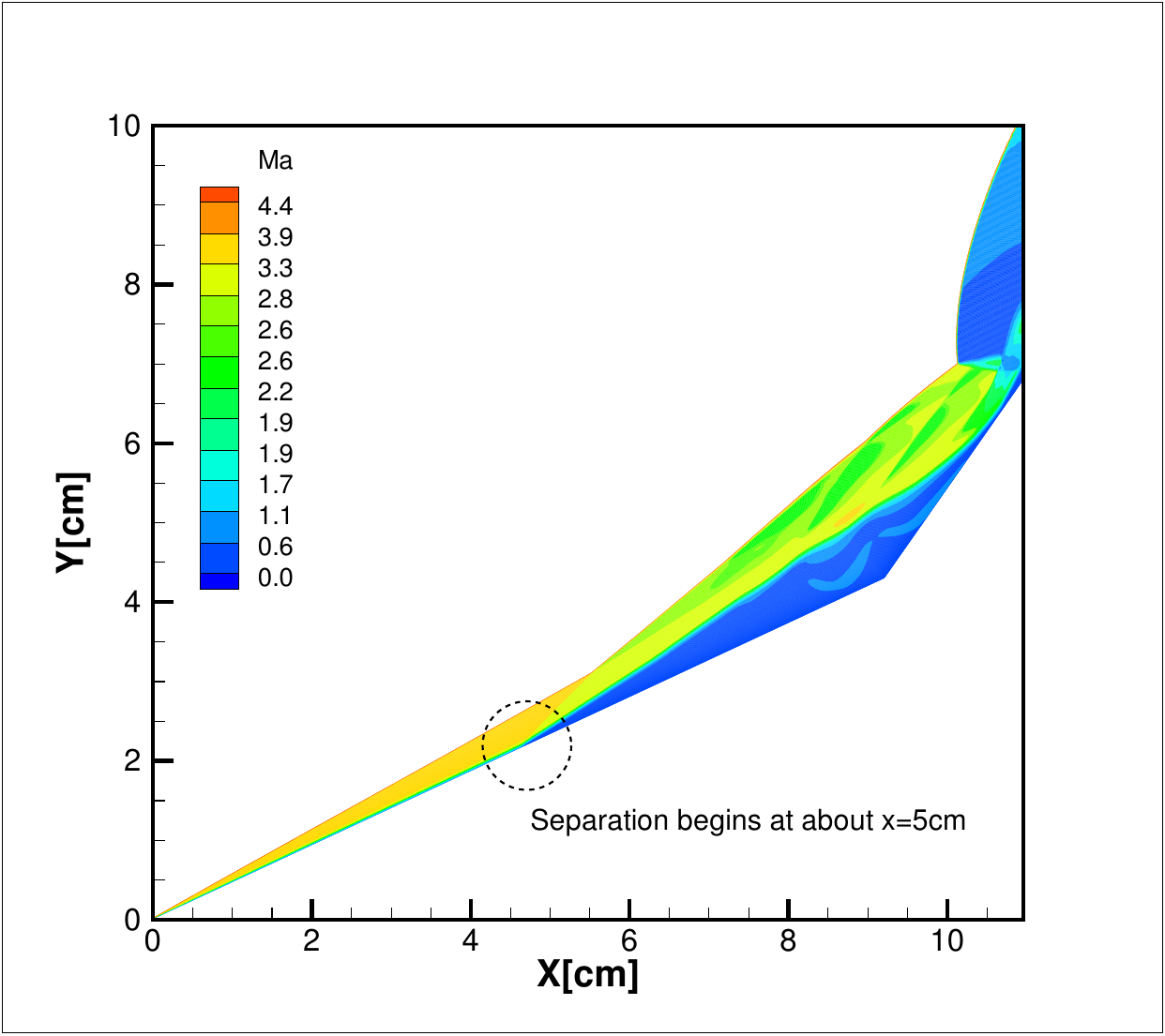}} 
	\caption{The Run46 case.  (\textit{a}) Pressure on the surface from the original (dashed line) and corrected (solid line) Eucken factors. (\textit{b}) Velocity profile along the normal direction of the surface at $x=2$~cm.  (\textit{c}) and (\textit{d}) Contours of the Mach number obtained from the original and corrected Eucken factors, respectively.}
	\label{fig:Run46_bijiao}
\end{figure}

The pressure loads along the double cone surface are shown in figure \ref{fig:Run46_bijiao}(\textit{a}). Like the heat flux, the location of separation moves upstream in our model of thermal conductivities, which matches the experiment data. Meanwhile, the pressure peak moves a little bit downstream. 
The adverse pressure gradient does not show clear difference. 

The velocity profile along the normal direction of the cone surface at $x=2$~cm is shown in figure \ref{fig:Run46_bijiao}(\textit{b}). The velocity profile from the corrected Eucken factor~\eqref{ft_fint_fv} is much smaller than that from the original one in \eqref{Euken_2}. This makes the flow separation from the corrected Eucken factor moves upstream, as evidenced by the abrupt drop in figure \ref{fig:Run2835804246surfaceheatflux}(\textit{d}). To be specific, figure \ref{fig:Run46_bijiao}(\textit{c}) shows that the separation position from the original Eucken factor locates at $x\approx5$~cm, while that from the corrected Eucken factor locates $x\approx6$~cm, see figure \ref{fig:Run46_bijiao}(\textit{d}). 

\begin{figure}
	\centering
	\subfigure[]{\includegraphics[viewport=20 10 540 500,clip=true,width=0.4\textwidth]{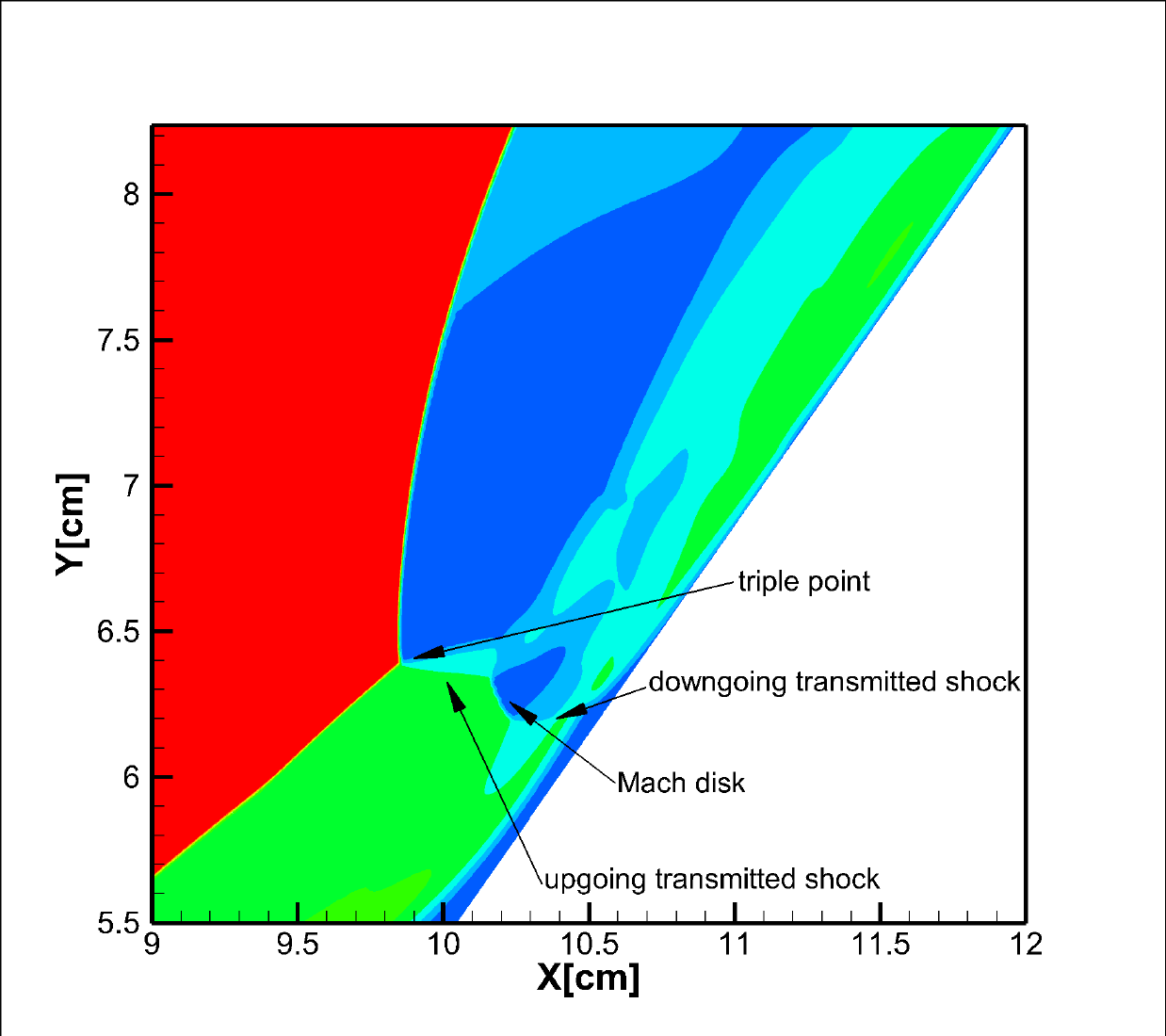}}
	\subfigure[]{\includegraphics[viewport=20 10 540 500,clip=true,width=0.4\textwidth]{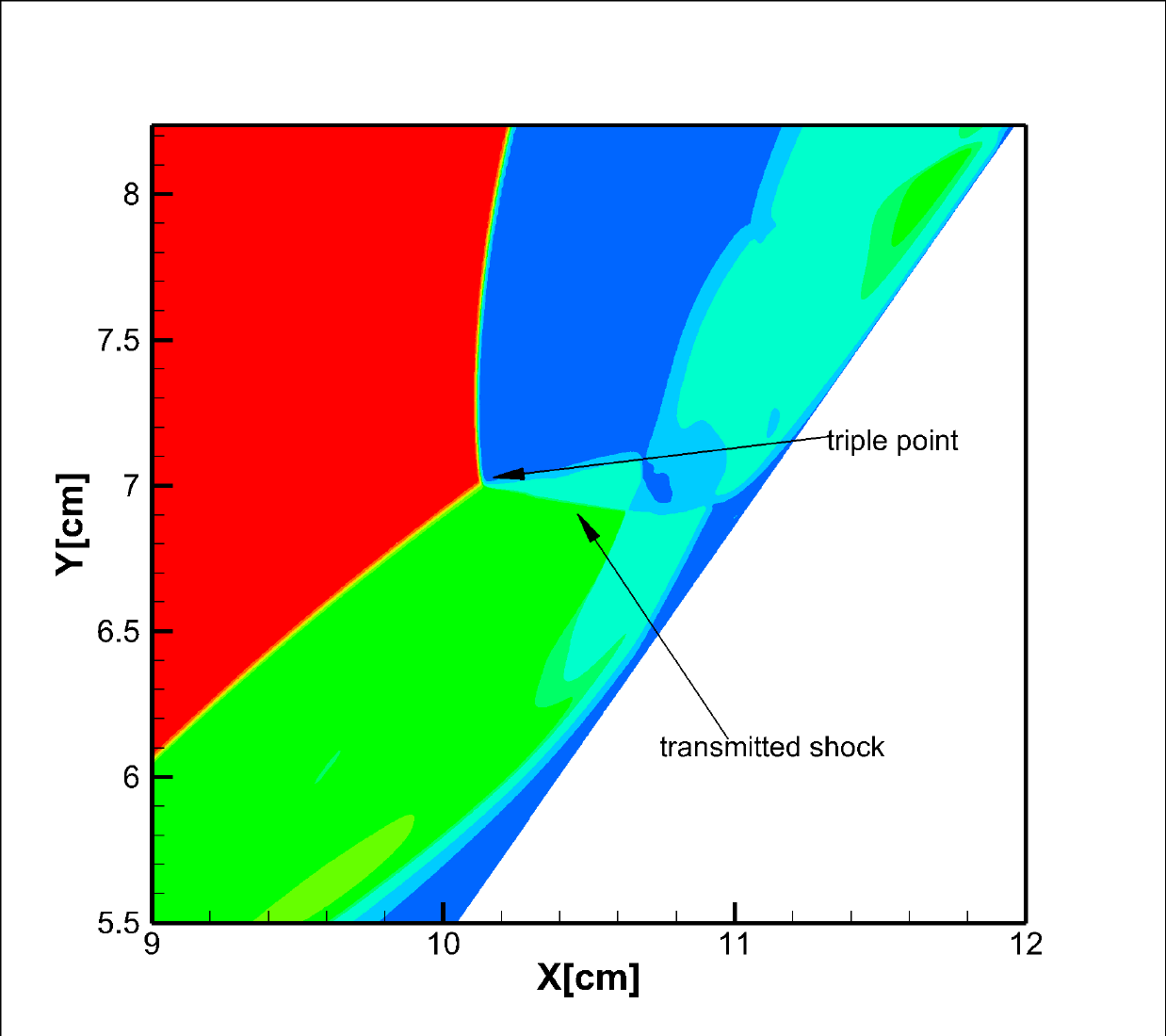}} 
	\caption{The Run46 case. Contours of the Mach number near the triple point, obtained from the original (\textit{a}) and corrected (\textit{b}) Eucken factors, respectively.}
	\label{fig:Run46_bijiao_2}
\end{figure}

The contours of the Mach number around the triple point are shown in figure \ref{fig:Run46_bijiao_2}(\textit{a}) and (\textit{b}) for the original and corrected Eucken factors, respectively. When the original Eucken factor is used in the NSF equations, the flow structure is quite similar to low enthalpy cases (e.g. Run28), which contains an upgoing transmitted shock, Mach disk and downgoing transmitted shock. The downgoing transmitted shock finally impacts upon the surface resulting in aerothermodynamic peak. However, when the corrected Eucken factor is adopted, the flow structure becomes quite simple which only contains a single transmitted shock from the triple point. This is quite different from the low enthalpy cases, where the modification of the Eucken factors rarely changes the flow pattern, see figure \ref{fig:Run28separatedshockbijiao}(e) and (f). As a result, the single transmitted shock impacts on the surface directly which makes the heat flux peak to a larger value than that from the original Eucken factor, see figure \ref{fig:Run2835804246surfaceheatflux}(\textit{d}).

\section{Influence of temperature jump conditions}\label{sec:TJCdiscussion}

To reveal the role of temperature jump conditions,  a comparison between the current method  \eqref{eq:temperature_jump_current} and the boundary condition without collision numbers is made.  The latter boundary condition is given by
\begin{equation}\label{translationalslip_withoutcn}
	\begin{aligned}
T_{t,j}=T_{w}+\frac{2-\sigma_t}{\sigma_t}\lambda \nabla T_t \cdot \boldsymbol{n},\\
	T_{r,j}=T_{w}+\frac{2-\sigma_r}{\sigma_r}\lambda \nabla T_r \cdot \boldsymbol{n},\\
		T_{v,j}=T_{w}+\frac{2-\sigma_v}{\sigma_v}\lambda \nabla T_v \cdot \boldsymbol{n},
\end{aligned}
\end{equation}
where $\sigma_t$, $\sigma_r$,  and $\sigma_v$ are the translational, rotation, and vibrational energy accommodation coefficients, respectively. To make a fair comparison, the same Eucken factors~\eqref{ft_fint_fv} are used.


It should be noted that \eqref{eq:temperature_jump_current} is founded on the assumption that during the gas-wall interactions, the internal energy of a gas molecule does not approach to that of the surface directly, but through the translational-internal energy relaxation of the gas. To be specific, the gas molecules need to go through rotational-translational and vibrational-translational relaxations first. Therefore, the rotational and vibrational energy accommodation processes are slower than that of the translational energy by about $Z_r$ and $Z_v$ times, respectively. Also 
when considering the interaction between nitrogen and stainless steel model, there is which makes $\frac{2-\sigma_{v}}{\sigma_{v}}=2000$ close to the order of $Z_v$.

\begin{figure}
  \centering
  \subfigure[]{\includegraphics[viewport=20 10 540 500,clip=true,width=0.3\textwidth]{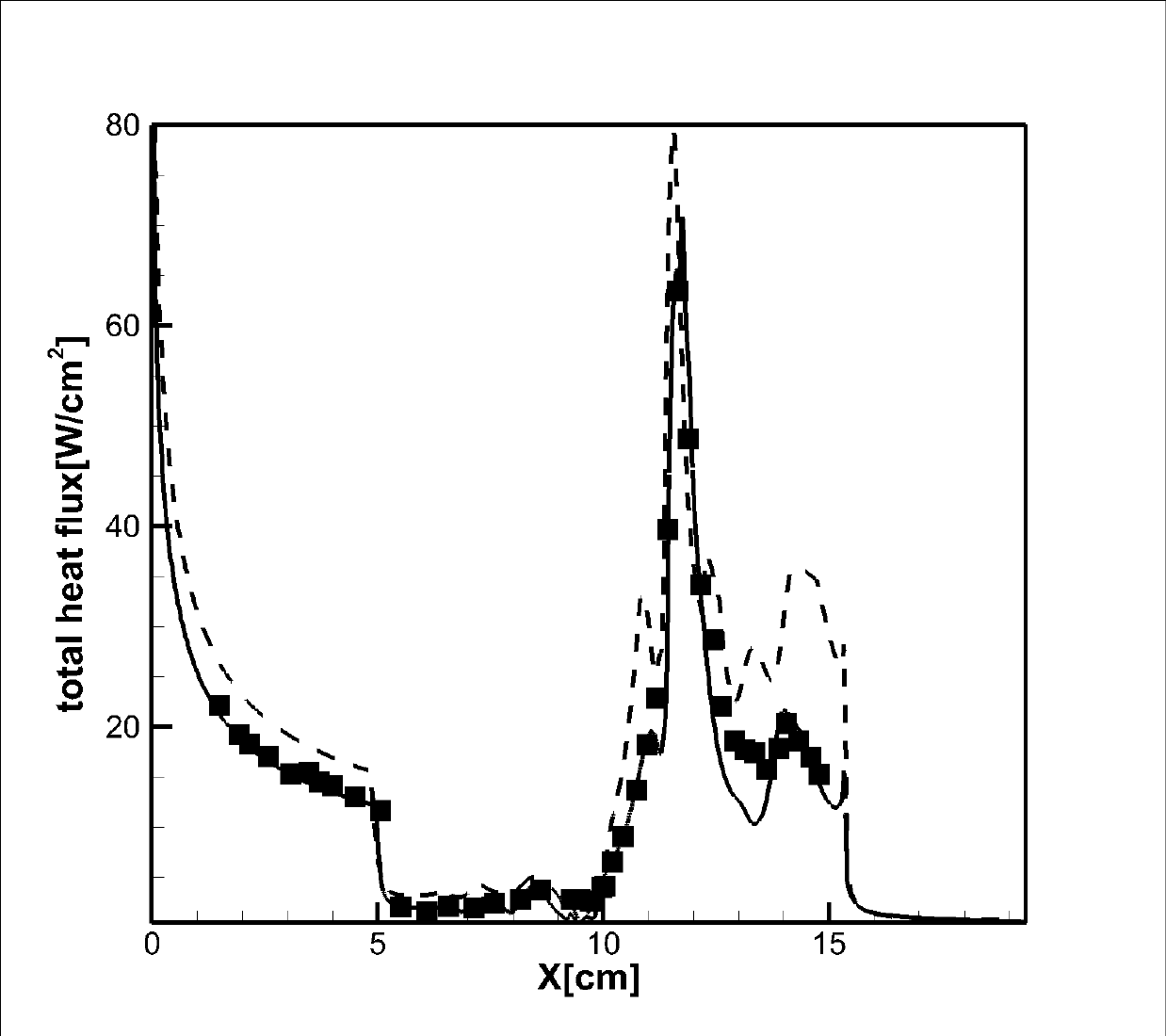}}
  \subfigure[]{\includegraphics[viewport=20 10 540 500,clip=true,width=0.3\textwidth]{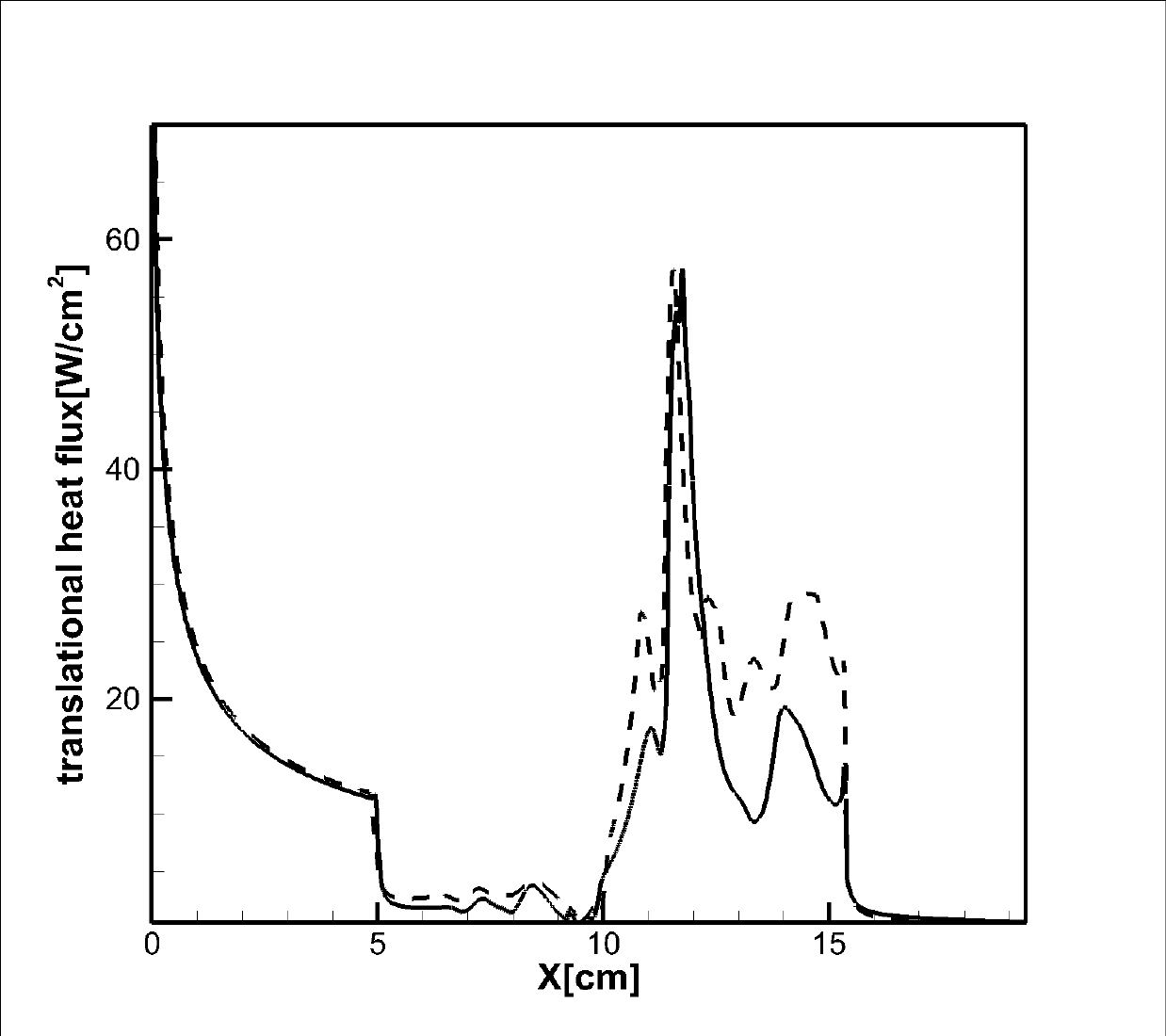}}
  \subfigure[]{\includegraphics[viewport=20 10 540 500,clip=true,width=0.3\textwidth]{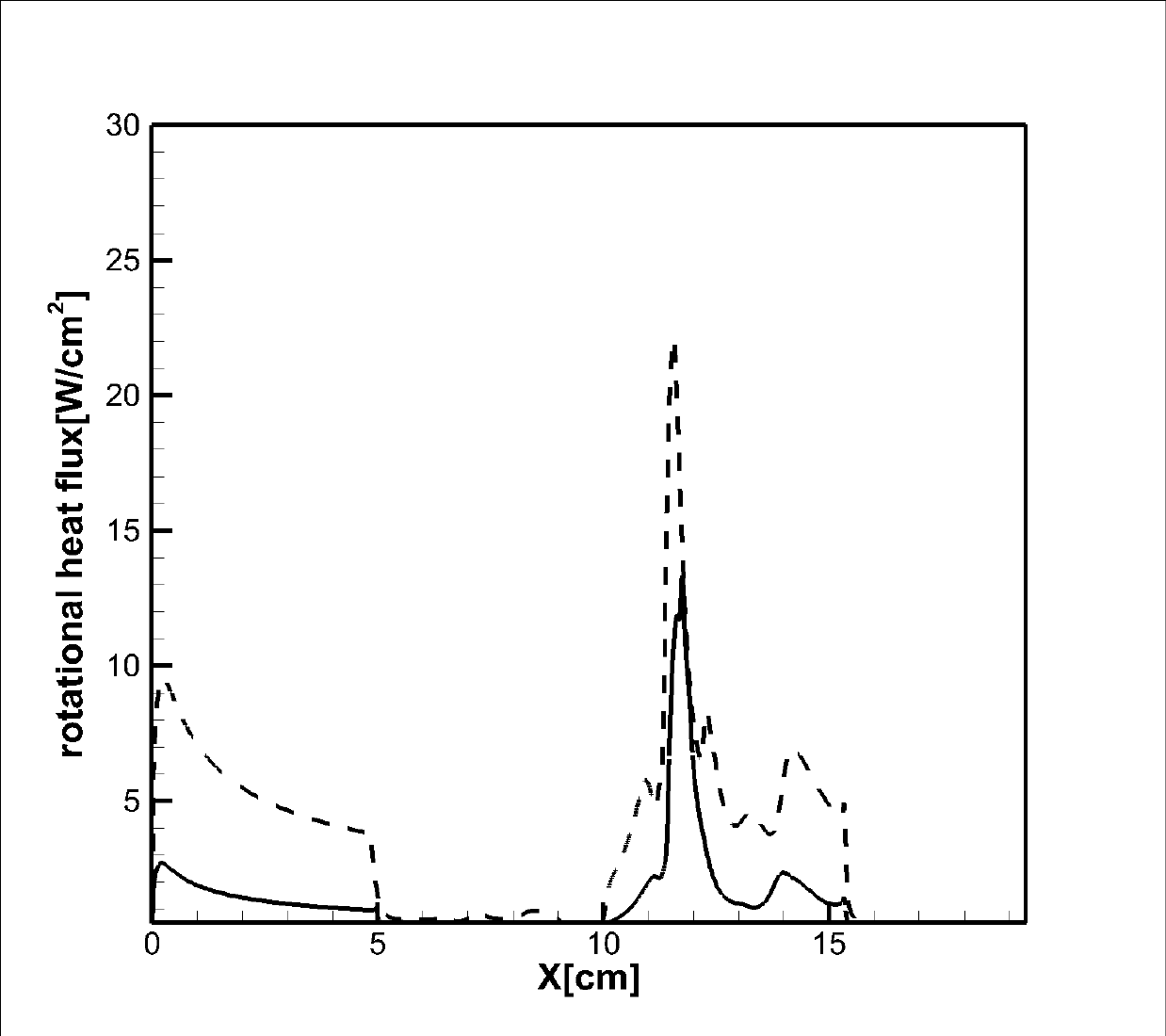}}
  \subfigure[]{\includegraphics[viewport=20 10 540 500,clip=true,width=0.3\textwidth]{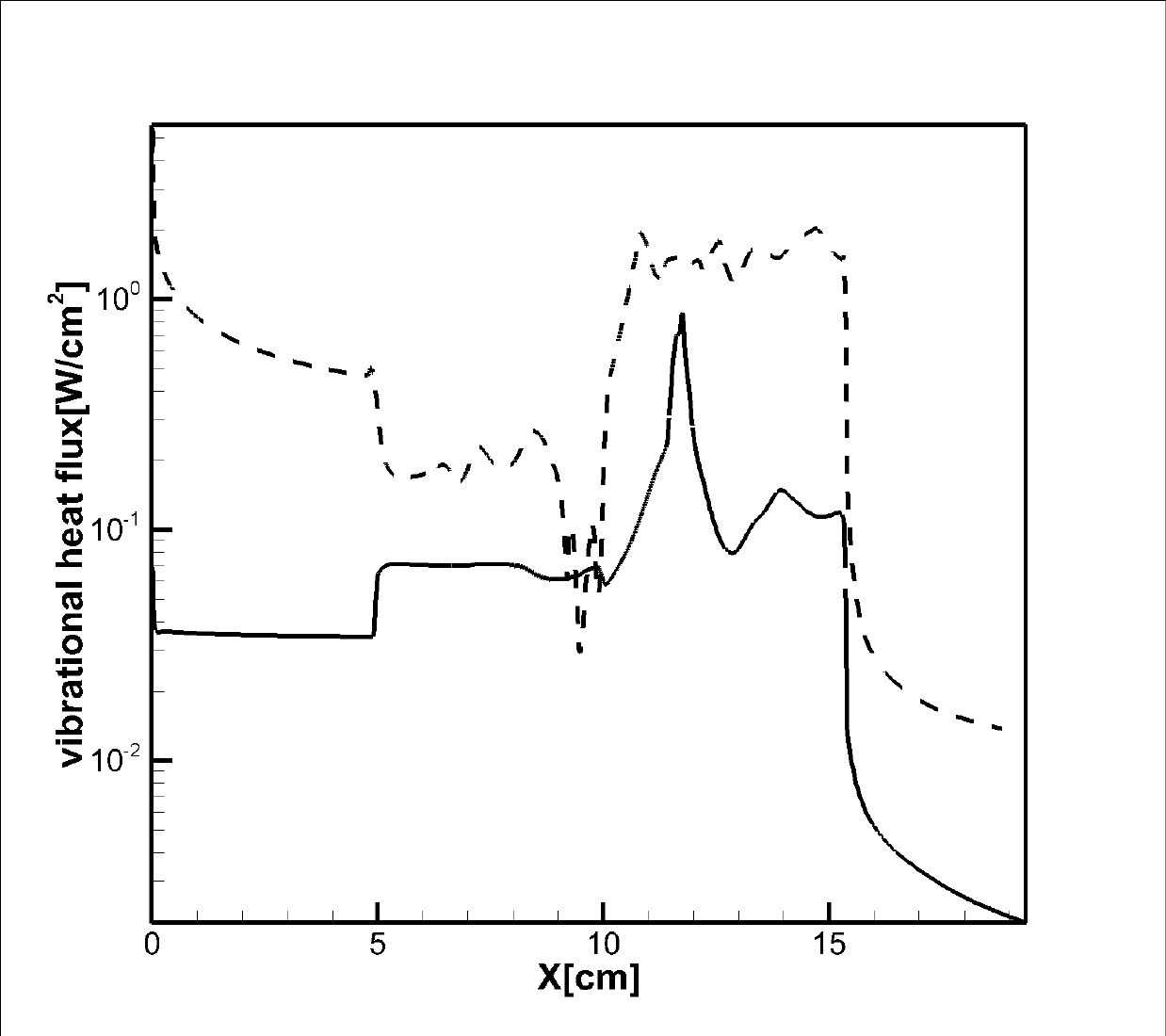}}
  \subfigure[]{\includegraphics[viewport=20 10 540 500,clip=true,width=0.3\textwidth]{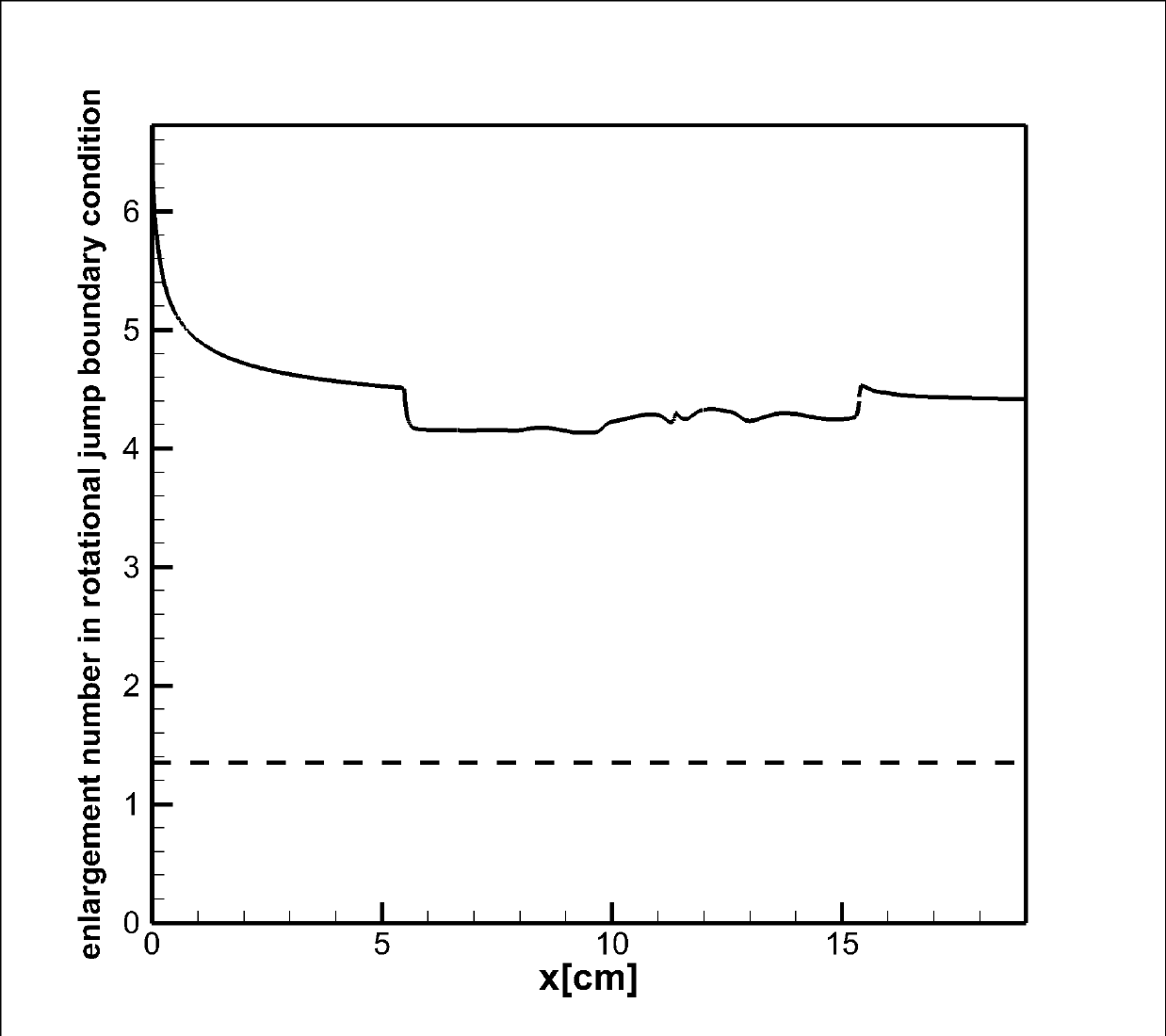}}
  \subfigure[]{\includegraphics[viewport=20 10 540 500,clip=true,width=0.3\textwidth]{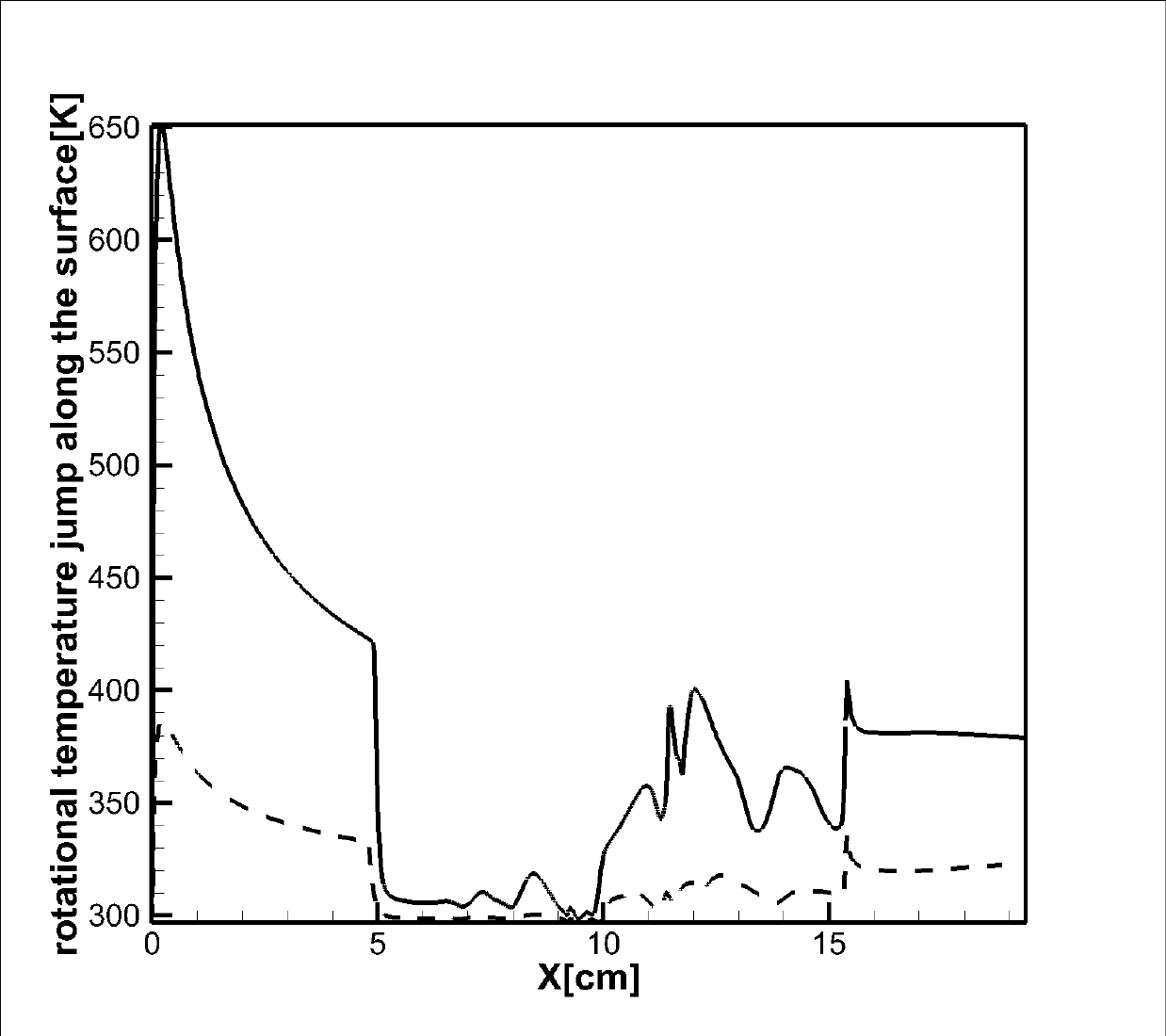}}
  \caption{Comparison of Run28 surface heat flux between different boundary condition: (\textit{a}) total heat flux, (\textit{b}) translational heat flux. (\textit{c}) rotational heat flux, (\textit{d}) vibrational heat flux. (\textit{e}) The rotational temperature jump coefficient $Z_r(2-\sigma)/\sigma=Z_r$ for the solid line, and 1.3529 for the dashed line used by \cite{Nompelis2004}.  (\textit{f}) Rotational temperature jump along the surface. Solid lines represent data from \eqref{eq:temperature_jump_current}, dashed lines represent data from \eqref{translationalslip_withoutcn}. Squares: experiment data.}
\label{fig:Run28slipbijiao}
\end{figure}

The heat flux of the Run28 case calculated from different boundary conditions are compared in figure \ref{fig:Run28slipbijiao}. A very good agreement with the experiment is obtained by using the new temperature-jump boundary condition \eqref{eq:temperature_jump_current}, where the jump coefficients, as given by
$Z_r(2-\sigma)/\sigma$ and  $Z_v(2-\sigma)/\sigma$,
are functions of the temperature.  
For further investigation, the translational, rotational and vibrational heat fluxes along the cone surface are plotted in figure \ref{fig:Run28slipbijiao}(\textit{b}), (\textit{c}) and (\textit{d}), respectively. The translational heat flux are nearly the same in figure \ref{fig:Run28slipbijiao}(\textit{b}), since the jump coefficient is not changed. Main difference is coming from rotational heat flux. As for the vibrational heat flux, due to the large temperature jump, its magnitude remains much less than the translational and rotational heat fluxes. So the difference in vibrational heat flux is neglected in current research. 

For further analyse into the difference in rotational heat flux, the rotational collision number in \eqref{eq:temperature_jump_current} and $\frac{2-\sigma_r}{\sigma_r}$ with $\sigma_r=0.85$ \citep{Nompelis2004} are plotted in figure \ref{fig:Run28slipbijiao}(\textit{e}). To be noticed  that $\frac{2-\sigma_r}{\sigma_r}$ is acting as the same role as $Z_r$. As a function of temperature, the rotational collision number $Z_r$ is three to four times of the constant $\frac{2-\sigma_r}{\sigma_r}$. This leads to the larger rotational temperature jump shown in figure \ref{fig:Run28slipbijiao}(\textit{f}), which leads to the smaller rotational heat flux from \eqref{eq:temperature_jump_current} shown in figure \ref{fig:Run28slipbijiao}(\textit{c}). Finally a smaller total heat flux is shown in figure \ref{fig:Run28slipbijiao}(\textit{a}), when compared to the temperature jump condition in \eqref{translationalslip_withoutcn}. 

It should be noticed that the difference in boundary conditions only change the magnitude of the heat flux, while the locations of abrupt drop and peak remain almost unaltered. Also, boundary condition do not change the location of separation bubble and reattachment point.

\section{Conclusions}

In summary, we have derived the three-temperature NSF equations from the modified Rykov kinetic equations, where the translational, rotational and vibrational Eucken factors are functions of the temperature. We have also proposed a new temperature-jump boundary conditions, where the jump coefficient is proportional to the rotational and vibrational collision numbers. With the two modifications, we have simulated both low- and high-enthalpy SWBLIs in the double cone. The numerical results have shown good agreement with experimental data in calculating the heat flux (including the drop, position and peak of the heat flux) and pressure along the cone surface. We have found that, the corrected Eucken factors could lead to a more accurate flow structure especially for the location of separation and the triple point, while the modified temperature-jump boundary condition only reduces the rotational heat flux so that the total heat flux agree with experimental measurement, but has no influence on the flow structure.

We have found that, compared to the original (widely-used) Eucken factor~\eqref{Euken_2}, the role of the modified Eucken factor~\eqref{ft_fint_fv} is different between the low- and high-enthalpy flows.
For low enthalpy cases, the corrected thermal conductivities \eqref{ft_fint_fv} predict a smaller adverse pressure gradient than the original heat conductivity model, but a similar velocity profile comparing to the original Eucken factor on the first cone. This leads to a delay in the boundary layer separation, and consequently a delay of steep drop of surface heat flux is predicted. As a response to this, the separated shock wave also starts later, leading to a different interaction position between the detached bow shock. After the triple point, a transmitted shock wave is generated. And the downgoing transmitted shock finally impacts on the surface, leading to reattachment. Different position of the triple point finally leads to different reattachment point. As a  consequence, a different peak in the surface heat flux is obtained. For high enthalpy cases, on the first cone, the corrected thermal conductivities \eqref{ft_fint_fv}  predict a similar adverse pressure gradient and smaller velocity profile  comparing to original Eucken factors \eqref{Euken_2}. This leads to an early boundary layer separation. The flow structures after the triple point are also different. Especially, a single transmitted shock wave rather than upgoing transmitted shock, Mach disk and downgoing transmitted shock is captured by using \eqref{ft_fint_fv}. The single transmitted shock wave directly impacts on the surface leading to a higher heat flux peak value, see figure \ref{fig:Run2835804246surfaceheatflux}(\textit{d}).

Finally, it is noted that, although we have only considered the single species nitrogen gas, our work sheds new light on improving the accuracy of high-temperature gas dynamic equations for multi-species gas mixtures, by investigating the components of heat conductivities and the temperature-jump conditions. 


\section{Acknowledgements}\label{sec:Acknowledgements}

This work is supported by the National Natural Science Foundation of China (12172162). Numerical simulations are conducted in the Centre for Computational Science and Engineering at the Southern University of Science and Technology.

\section{Declaration of interest}\label{sec:intrest}

The authors report no conflict of interest.

\bibliographystyle{jfm}
\bibliography{jfm}



\end{document}